\documentclass[envcountsect,envcountsame,leqno]{llncs}

\usepackage{amsmath,amssymb,euscript}
\usepackage{graphicx}
\usepackage{comment}
\usepackage{xspace}
\usepackage{multirow}
\usepackage{url}
\usepackage{cite}
\usepackage{rotating}
\usepackage[shortlabels]{enumitem}
\usepackage{color}
\usepackage{subfigure}

\def\scri#1{{\EuScript#1}}

\newcommand{\crg}{\mathrm{cr}}
\newcommand{\crgapx}{\mathrm{cr}^\mathrm{aprx}}
\newcommand{\ins}{\mathrm{ins}}
\newcommand{\inslb}{\mathrm{ins}^\Sigma}

\newcommand{\insapx}{\mathrm{ins}^\mathrm{aprx}}
\def\MEI(#1,#2){\mathrm{MEI}(#1,#2)}

\newcommand{\lswitching}{\textsc{switching}\xspace}
\newcommand{\lnonswitching}{\textsc{nonswitching}\xspace}

\newcommand{\lflipped}{\textsc{flipped}\xspace}
\newcommand{\lnonflipped}{\textsc{nonflipped}\xspace}

\newcommand{\defend}{\hfill$\diamond$}

\spnewtheorem{defn}[theorem]{Definition}{\bf}{\rm}
\spnewtheorem{claimx}[theorem]{Claim}{\bf}{\it}

\spnewtheorem{obs}[theorem]{Observation}{\bf}{\it}
\spnewtheorem{alg}[theorem]{Algorithm}{\bf}{\rm}

\begin{document}
\title{A Tighter Insertion-based Approximation\\of the Crossing Number\thanks{A preliminary version of this research appeared in ICALP 2011~\cite{CHicalp}.}}
\author{Markus Chimani\inst{1}\and 
	Petr Hlin\v{e}n\'y\thanks{P.~Hlin\v{e}n\'y has been supported by
	 the Czech Science Foundation project 14-03501S.}\inst{2}}
\institute{Theoretical Computer Science, University Osnabr\"uck, Germany\\
\email{markus.chimani@uni-osnabrueck.de} \and
Faculty of Informatics, Masaryk University Brno, Czech Republic\\
\email{hlineny@fi.muni.cz}\\[2ex]
\today
}
\maketitle

\begin{abstract}
Let $G$ be a planar graph and $F$ a set of additional edges not yet in $G$. The
{\em multiple edge insertion} problem (MEI) asks for a drawing of $G+F$ with the
minimum number of pairwise edge crossings, such that the subdrawing of $G$ is
plane. Finding an exact solution to MEI is NP-hard for general $F$.
We present the first polynomial time algorithm for MEI that achieves an additive 
approximation guarantee -- depending only on the size of $F$ and the maximum degree of
$G$, in the case of connected~$G$. 
Our algorithm seems to be the first directly implementable one in that realm, too,
next to the single edge insertion.

It is also known that an (even approximate) solution to the MEI problem would
approximate the crossing number of the \emph{$F$-almost-planar graph} $G+F$,
while computing the crossing number of $G+F$ exactly is NP-hard already when
$|F|=1$. Hence our algorithm induces new, improved approximation bounds for the
crossing number problem of $F$-almost-planar graphs, achieving constant-factor
approximation for the large class of such graphs of bounded degrees and bounded
size of $F$.
\end{abstract}

\section{Introduction}

The crossing number $\crg(G)$ of a graph $G$ is the minimum number of pairwise
edge crossings in a drawing of $G$ in the plane. Finding the cossing number of a graph
is one of the most prominent combinatorial optimization problems in 
graph theory and is NP-hard already in very restricted cases, e.g.,
even when considering a planar graph with one added edge~\cite{CM-socg2010}.
It has been vividly investigated for over 60 years, but there is still surprisingly
little known about it; 
see~\cite{vrto} for an extensive bibliography. 
While the approximability status of the crossing number problem is still unknown, 
several approximation algorithms arose for special graph classes. 

For the crossing number of general graphs with bounded degree, there is an algorithm~\cite{DBLP:journals/jcss/BhattL84} that
approximates not directly the crossing number, but the quantity $n+\crg(G)$; 
its currently best incarnation does so within a factor of $O(\log^2 n)$~\cite{egs}. 
In terms of pure $\crg(G)$-approximation this hence resembles a ratio of $O(n\log^2 n)$. The first sublinear approximation 
factor of $\tilde O(n^{0.9})$ was recently given via a highly involved (and unfortunately not directly practical) algorithm~\cite{stoccrossing}.

The known constant factor
approximations restrict themselves to graphs following one of two paradigms (see
also Section~\ref{sec:crApprox}): they either assume that the graph is
embeddable in some higher surface~\cite{ghls,torusHS,surfaceApprox}, or they are
based on the idea that only a small set of graph elements has to be removed from
$G$ to make it planar: removing and re-inserting them can give strong
approximation bounds \cite{HS06,cm08,apex}. In this paper, we follow the latter idea
and first concentrate on the following tightly related problem:

\begin{defn}[Multiple edge insertion, MEI]\label{def:MEI}
Let $G$ be a planar graph and $F$ a set of edges (vertex pairs, in fact) 
not in~$E(G)$. We denote by $G+F$ the graph obtained by adding $F$ to the edge
set of~$G$.
The \emph{multiple edge insertion} problem $\MEI(G,F)$ is to find
\begin{enumerate}[i)]
\item
a plane embedding $G_0$ of $G$, and
\item
a drawing $G_F$ of the graph $G+F$ such that the restriction of $G_F$ to $G$
is the plane embedding $G_0$,
\end{enumerate}
such that the number of pairwise edge crossings in $G_F$ is minimized over
all plane embeddings $G_0$ of $G$ and all drawings $G_F$ as in (ii).

Let $\ins(G_0,F)$ denote the minimum number of edge crossings of $G_F$ as 
in~(ii) and let $\ins(G,F)$, the {\em solution size} of the $\MEI(G,F)$ problem,
denote this minimum of $\ins(G_0,F)$ over all plane embeddings $G_0$ of $G$.
\defend\end{defn}

We refer to Section~\ref{sec:prelim} for formal definitions of a drawing and
edge crossings.
Observe that, in a solution to MEI, each crossing hence involves at least one edge of $F$.
This can be a severe restriction, as shown by the examples of $\MEI(G,F)$ instances
with solution sizes much larger than $\crg(G+F)$ given, e.g.,
in~\cite{GMW05,HS06}.

For general $k=|F|$, the MEI problem is known to be NP-hard~\cite{zieglerTh}, based
on a reduction from \emph{fixed linear crossing number} (see Appendix); 
for fixed $k>1$ the problem complexity is open.
The main difficulty of the MEI problem, roughly, comes from possible
existence of (up to exponentially many) inequivalent embeddings of~$G$.

The case $k=1$ of MEI is known as the \emph{(single) edge insertion} problem and
can be solved optimally in linear time~\cite{GMW05}, as we will briefly summarize in
Section~\ref{sec:singleIns}. 
Let $e$ be the edge to insert, and denote the resulting
number of crossings by $\ins(G,e)$.  
Let $\Delta(G)$ denote the maximum degree in $G$.
It was shown \cite{HS06,cm08} that
$\ins(G,e)$ approximates the crossing number $\crg(G+e)$---i.e., of the graph
containing this edge~$e$---within a multiplicative factor of 
$\lfloor\frac12\Delta(G)\rfloor$ achieved in \cite{cm08}, and this bound is tight.
Recall also that computing $\crg(G+e)$ exactly is
NP-hard~\cite{CM-socg2010}.

Another special case of the MEI problem is when one adds a new vertex together with its
incident edges; this is also polynomially solvable~\cite{insertvertex} and
approximates the crossing number of the resulting \emph{apex graph}~\cite{apex}.  
A slight variant of this \emph{vertex insertion} (or \emph{star insertion}) problem, is the \emph{multiple adjacent edge insertion},
were the inserted edges have a common incident vertex that need not be new. 
The algorithm in~\cite{insertvertex} also solves this latter variant.
However, these are the only types of insertion problems that are currently known to be in~P.  

Nevertheless, it has been proven in \cite{apex} (see Section~\ref{sec:crApprox}) 
that a solution (even an approximate one) to $\MEI(G,F)$
would directly imply an approximation algorithm for $\crg(G+F)$ with planar $G$.
Independently, Chuzhoy et al.~\cite{sodamultiedge} have shown
the first algorithm efficiently computing an approximate
solution to the crossing number problem on $G+F$ with the help of
a multiple edge insertion solution. 
Precisely, they have achieved a solution with the number of crossings
\begin{align}
\crgapx(G+F) \>\leq\> O\big(\Delta(G)^3\cdot|F|\cdot\crg(G+F)+\Delta(G)^3\cdot|F|^2\big)
\label{eq:chuzhoy}
\end{align}
(without giving explicit constants).
Though not mentioned explicitly in \cite{sodamultiedge},
it seems that their results also give an approximation solution to $\MEI(G,F)$ with the same ratio,
at least in the case of $3$-connected $G+F$.
However, the algorithm \cite{sodamultiedge} is unfortunately not directly applicable in practice.

In this paper, we pay our main attention to the MEI problem.
In contrast to a multiplicative factor as in~\eqref{eq:chuzhoy}, we
provide an efficient algorithm approximating a solution of $\MEI(G,F)$ with an
additive approximation guarantee~\eqref{eq:ourMEI}.
Then we employ the aforementioned generic result of \cite{apex} to derive a
corresponding approximation of the crossing number (this up to a multiplicative factor).
On the one hand, our approach is algorithmically and implementationally simpler, virtually
only building on top of well-studied and experimentally evaluated sub-algorithms.
On the other hand, it gives stronger approximations also for the crossing
number, cf.~\eqref{eq:ourCR} as compared to~\eqref{eq:chuzhoy}, as well as better runtime bounds.

We are going to show:

\begin{thm}\label{thm:main}
Given a connected planar graph $G$ and an edge set $F$, $F\cap E(G)=\emptyset$.
Let $k:=|F|$ and $\Delta:=\Delta(G)$. 
Algorithm~\ref{alg:MEIapx} described below finds, in 
$O(k\cdot|V(G)|+k^2)$ time, a~solution to the $\MEI(G,F)$ problem 
with $\insapx(G,F)$ crossings such that
\begin{align}
\insapx(G,F) &= \ins(G,F)+ O(\Delta\, k\log k + k^2)\mbox{, or more precisely}\\
\insapx(G,F) &\leq \ins(G,F)+ 
	 2k\,\lfloor\log_2k\rfloor\cdot\left\lfloor\mbox{$\frac12$}\Delta\right\rfloor
	+\mbox{$\frac12$}\big(k^2-k\big)\label{eq:ourMEI} .
\end{align}
Consequently, this gives an approximate solution to the crossing number
problem
\begin{align}
\crgapx(G+F) &= O(\Delta k)\cdot\crg(G+F)
	+ O(\Delta\, k\log k + k^2)\mbox{, or more precisely}\\
\crgapx(G+F) &\leq \lfloor\mbox{$\frac12$}\Delta\rfloor\!\cdot 2k\!\cdot\crg(G+F)
	+ 2k\,\lfloor\log_2k\rfloor\cdot\left\lfloor\mbox{$\frac12$}\Delta\right\rfloor
	+\mbox{$\frac12$}\big(k^2-k\big)\label{eq:ourCR} .
\end{align}
\end{thm}

Notice the constant-factor approximation ratio when the degree of $G$ and the size
of $F$ are bounded.
We remark that the assumption of connectivity of $G$ is
necessary in the context of Theorem~\ref{thm:main}. For disconnected $G$,
the approximation guarantee for $\insapx(G,F)$ would be the same as for
$\crgapx(G+F)$ in \eqref{eq:ourCR}.

Concerning practical usability,
our Algorithm~\ref{alg:MEIapx}, in fact, seems to be the first {\em directly implementable}
and practically useful algorithm in this area, next to the single edge insertion. 
In \cite{DBLP:journals/jgaa/ChimaniG12}, an 
implementation of this algorithm is compared to the strongest known heuristics in practice, and offers the
arguably best balance between running time (being faster than most heuristics) and solution quality (second only to a very long running heuristic).

\section{Decompositions and Embedding Preferences}\label{sec:prelim}

We use the standard terminology of graph theory.
By default, we use the term {\em graph} to refer to a {loopless multigraph}.
This means that we allow {\em parallel edges} (but no loops),
and when speaking about a {\em cycle} in a graph, we include also the case
of a $2$-cycle formed by a pair of parallel edges.
If there is no danger of confusion between parallel edges, we denote 
an edge with the ends $u$ and $v$ chiefly by $uv$.

We pay particular attention to graph connectivity.
A {\em cut vertex} in a connected graph $G$ is a vertex $u$ such that 
$G-u$ is disconnected.
A graph $G$ is {\em biconnected} if $G$ has at least $3$ vertices and no
cut vertex,
and $G$ is {\em triconnected} if $G$ has at least $4$ vertices and no
vertex-cut of size $\leq2$.
A {\em block} in a graph $G$ is a maximal subgraph $H\subseteq G$ such that
$H$ contains no cut vertex (of $H$).
In other words, a block $H$ is a maximal biconnected subgraph of $G$,
or $H$ has at most $2$ vertices, which are not contained in any common larger block.

A \emph{drawing} of a graph $G=(V,E)$ is a mapping of the vertices $V$ to distinct points on a surface $\Sigma$,
and of the edges $E$ to simple curves on $\Sigma$, 
connecting their respective end points but not containing any other vertex point.
Unless explicitly specified, we will always assume $\Sigma$ to be the plane (or, equivalently, the sphere).
A \emph{crossing} is a common point of two distinct edge curves, other than their common end point.
Then, a drawing is \emph{plane} if there are no crossings. 
A graph is \emph{planar}, if and only if it allows a plane drawing.

The \emph{crossing number} problem asks for a drawing of a given graph $G$ with the least 
possible number $\crg(G)$ of pairwise edge crossings
(while there exists other definitions of a crossing number such as the pair
or odd crossing numbers, those are not the subject of our paper).
By saying ``pairwise edge crossings'' we would like to emphasize that we count
a crossing point $x$ separately for every pair of edges meeting in $x$
(e.g., if $k$ edges meet in $x$, then this accounts for ${k\choose2}$ crossings).
It is well established that the search for an optimal solution to the above crossing
number problem can be restricted to so called \emph{good} drawings:
any pair of edges crosses at most once, adjacent edges do not cross, and there is no point on $\Sigma$ 
that is a crossing of three or more edges.

A \emph{plane embedding} $G_0$ of a connected graph $G$ is $G$ augmented with the cyclic
orders of the edges around their incident vertices, such that there is a
plane drawing of $G$ respecting these orders. 
Embeddings hence form equivalence classes over all plane drawings of $G$ on the sphere.
Choosing any of the thereby induced \emph{faces}---the regions on the sphere enclosed by 
edge curves---as the (infinite, unbounded) \emph{outer face} gives an
actual drawing in the plane.
However, it is technically easier to keep the ``freedom of choice'' of the
outer face, and hence to work with plane drawings and embeddings on the sphere
(unless we explicitly specify otherwise).
Observe that, when inverting all the cyclic orders of an embedding, we 
\emph{mirror} the embedding, and consequently the corresponding drawings.
In regard of Definition~\ref{def:MEI}, if
$G_F$ is a drawing of the graph $G+F$ such that the restriction of $G_F$ to $G$
gives a plane embedding $G_0$, then we chiefly refer to $G_F$ as to $G_0+F$.

Given a plane embedding $G_0$ of $G$, we define its \emph{dual} $G_0^*$ as the embedded graph that has a (dual) vertex for each face in $G_0$;
dual vertices are joined by a (dual) edge for each (primal) edge shared by their respective (primal) faces. The cyclic order of the (dual) edges around
any common incident (dual) vertex~$v^*$, is induced by the cyclic order of the (primal) edges around the (primal) face corresponding to~$v^*$. 
We may refer to a path in $G_0^*$ as to a \emph{dual path} in $G_0$.

\subsection{Insertion problems and decomposition trees}
\label{sub:decomptree}

When dealing with insertion problems, we always consider a connected planar graph $G$ 
with a set $F$ of additional $k$ edges (with the ends in $V(G)$) not present in $E(G)$. 
Note that, since we allow multigraphs, an edge from $F$ may be parallel to an
existing edge of~$G$.

Insertion algorithms typically work in two phases: first, 
they choose an embedding $G_0$ of $G$; then they fix $G_0$ and draw the edges $F$ within it.
In this context, we also use %(mainly in algorithmic settings) 
the following terminology.
\begin{defn}[Insertion path]\label{def:insertionpath}
Consider a connected planar graph $G$ and $v_1,v_2\in V(G)$.
Let $G_0$ be a plane embedding of~$G$.
An {\em insertion path of $\{v_1,v_2\}$} is a shortest dual path in $G_0$
from a face incident to $v_1$ to a face incident to~$v_2$.%
\defend\end{defn}

\begin{claim}
Let $G_0$ be a plane embedding of a graph and $v_1,v_2\in V(G_0)$.
A new edge $v_1v_2$ can be drawn in $G_0$ with at most $k$ crossings
if, and only if, there is an insertion path of $\{v_1,v_2\}$ in $G_0$ of
length at most~$k$.
\qed\end{claim}

Our approach to edge insertion
will use suitable tree-structured decompositions of the given planar graph, 
according to its connectivity. 
The concept of these decompositions is also illustrated with an example
in Figure~\ref{fig:exampledecomp}.

\begin{defn}[BC-tree]
Let $G$ be a connected graph. The \emph{BC-tree} $\scri B=\scri B(G)$ 
of $G$ is a tree that satisfies the following properties: 
\begin{enumerate}[i)]
\item $\scri B$ has two different node types: B- and C-nodes.
\item For every cut vertex in $G$,~ $\scri B$ contains a unique corresponding C-node.
\item For every \emph{block} in $G$,~ $\scri B$ contains a unique corresponding B-node.
\item No two B-, and no two C-nodes are adjacent. A B-node is adjacent to a 
C-node iff the corresponding block contains the corresponding cut vertex.\defend
\end{enumerate}
\end{defn}

To further decompose the blocks, we consider SPQR-trees for each non-trivial B-node 
(i.e., whose block contains more than two vertices).
This decomposition was first defined 
in~\cite{DT96}, based on prior work of~\cite{BM90,Tut66}. Even though more 
complicated than the 
BC-tree, it requires also only linear size and can be constructed in linear
time~\cite{HT73,GM01}. We are mainly interested in the property
that an SPQR-tree can be used to efficiently represent and enumerate all
(potentially exponentially many) plane embeddings of its underlying graph.
For conciseness, we call our tree \emph{SPR-tree}, as we do not require nodes of 
type Q.%

\begin{defn}[SPR-tree, cf.~\cite{mchDiss}]
\label{def:SPR}
Let $H$ be a biconnected graph with at least three vertices. 
The \emph{SPR-tree} $\scri T$ of $H$ is the (unique) smallest 
tree satisfying the following properties: 
\begin{enumerate}[i)]
\item Each node $\nu$ in $\scri T$ holds a specific (small) graph $S_\nu=(V_\nu,E_\nu)$
where $V_\nu\subseteq V(H)$, called a \emph{skeleton}.
Each edge $f$ of $E_\nu$ is either a {\em real} edge $f\in E(H)$,
or a {\em virtual} edge $f=uv\not\in E(H)$ (while still, $u,v\in V(H)$).
  \item $\scri T$ has three different node types with the following skeleton structures:
\begin{itemize}
\item[{\bf S:}] the skeleton $S_\nu$ is a cycle (of length $2$ or more)---it 
represents a \emph{serial} component; %There are no adjacent S-nodes in $\cal T$.
\item[{\bf P:}] the skeleton $S_\nu$ consists of two vertices and
at least three multiple edges between them---it represents a \emph{parallel} component;
\item[{\bf R:}] the skeleton $S_\nu$ is a simple triconnected graph
on at least four vertices---it is {\em``rigid''}.
\end{itemize}
\item\label{it:emuenu}
For every edge $\nu\mu$ in $\scri T$ we have $|V_\nu\cap V_\mu|=2$. 
These two common vertices, say $x,y$, form a vertex $2$-cut
(a {\em split pair}) in~$H$. 
Skeleton $S_\nu$ contains a specific virtual edge $e_\mu\in E(S_\nu)$ that 
represents the node $\mu$
and, symmetrically, some specific $e_\nu\in E(S_\mu)$ represents~$\nu$;
both $e_\nu,e_\mu$ have the ends $x,y$. These two virtual edges may refer to one another as \emph{twins}.
\item\label{it:gluev}
The original graph $H$ can be obtained by recursively applying the following 
operation of {merging}:
For an edge $\nu\mu\in E(\scri T)$, let $e_\mu$, $e_\nu$ be the twin pair of virtual
edges as in (\ref{it:emuenu} connecting the same $x,y$.
A {\em merged} graph \mbox{$(S_\nu\cup S_\mu)-\{e_\mu,e_\nu\}$}
is obtained by gluing the two skeletons together at $x,y$ and removing $e_\mu,e_\nu$.\defend
\end{enumerate}
\end{defn}

We remark that SPQR-trees have also been used in the aforementioned
\cite{sodamultiedge}, though with a different approach.
We use a slightly modified version of the SPR-tree,
which ``inserts'' a degenerate S-node between each pair of P- or R-nodes:

\begin{defn}[sSPR-tree]
\label{def:sSPR-tree}
Let $H$ be a biconnected graph with at least three vertices. 
A \emph{serialized SPR-tree} (\emph{sSPR} for short)
$\scri T=\scri T(H)$ of $H$ is the unique smallest
tree satisfying the properties (i)--(iv) of SPR-trees and additionally the following;
\begin{enumerate}[i)]
\setcounter{enumi}{4}%
\item every edge in $\scri T$ has precisely one end being an S-node.\defend
\end{enumerate}
\end{defn}

In traditional SPR-trees, S-node skeletons will always contain at least three edges, 
due to their minimality. 
Because of property (v), 
sSPR-trees may, however, now also contain S-nodes representing 2-cycles. 
In fact, it is trivial to obtain an sSPR-tree from
an SPR-tree by subdividing any edge that it not incident to an S-node with such a 2-cycle S-node. 
We observe that the sSPR-tree retains essentially
all properties of SPR-trees, in particular it also has only linear size, 
can be computed in linear time, and all the previously known insertion algorithms 
(most importantly the single edge insertion algorithm~\cite{GMW05}) 
can be performed using the sSPR-tree without any modifications.

\newcommand{\mybigscale}{0.6}
\newcommand{\myscale}{0.5}
\newcommand{\mysmallscale}{0.4}
\begin{figure}
\begin{minipage}{0.65\textwidth}
Given graph:\\
\includegraphics[scale=\mybigscale]{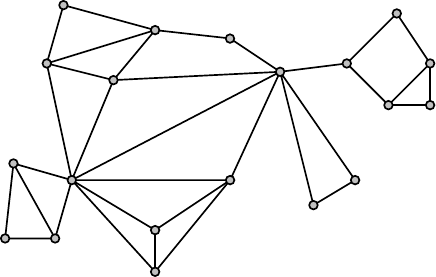}

\bigskip
BC-tree:\\
\includegraphics[scale=\myscale]{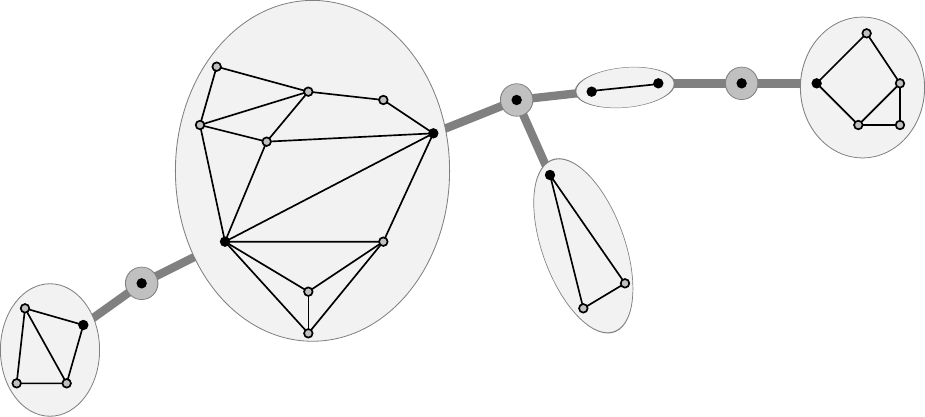}

\bigskip
Con-tree:\\
\includegraphics[scale=\myscale]{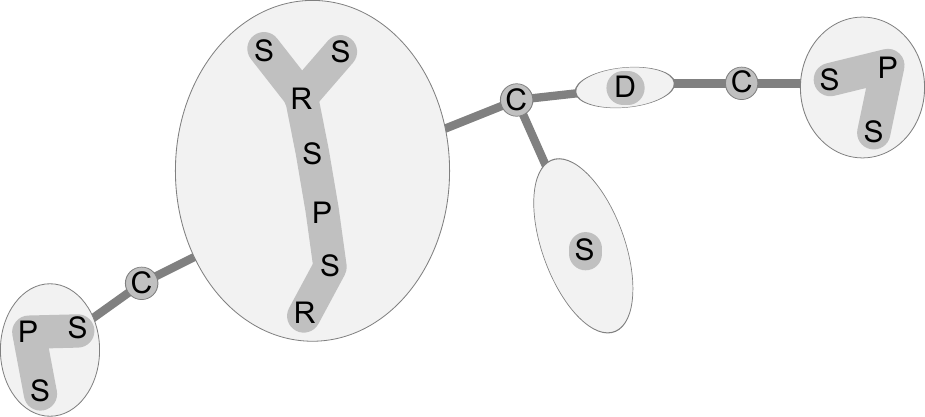}

\bigskip
Decomposition graph:\\
\includegraphics[scale=\myscale]{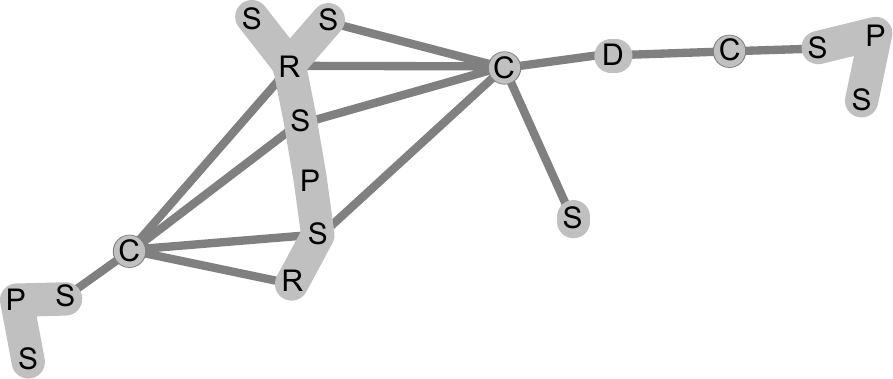}
\end{minipage}\hfill
\begin{minipage}{0.3\textwidth}
\raggedleft
SPR-tree \& skeletons\\of largest block:\\
\includegraphics[scale=\mysmallscale]{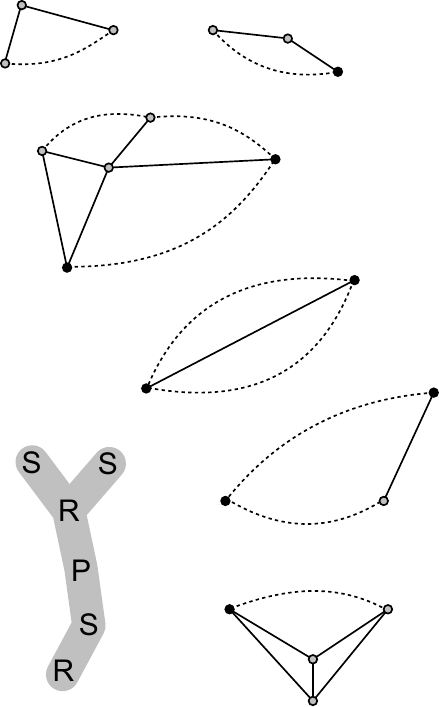}

\bigskip
\bigskip
sSPR-tree \& skeletons\\of largest block:\\
\includegraphics[scale=\mysmallscale]{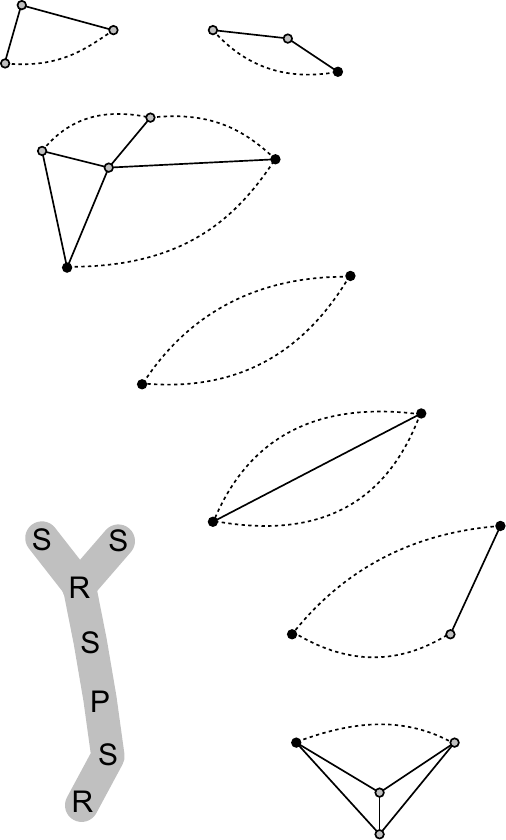}
\end{minipage}
\caption{Example for the various decompositions.}
\label{fig:exampledecomp}
\end{figure}

We are particularly interested in the amalgamated
version of the above decompositions, chiefly denoted by \emph{con-tree}:
\begin{defn}[Con-tree]
\label{def:con-tree}
Given a connected, planar graph $G$, let the {\em con-tree} $\scri C=\scri
C(G)$ be formed of the BC-tree $\scri B(G)$ 
that holds sSPR-trees $\scri T(H)$ for all non-trivial blocks $H$ of $G$. 
For technical reasons, if $H$ is a trivial $2$-vertex block, 
we set $\scri T(H)$ to be the tree formed by a single dummy
node, called a {\em D-node}, whose skeleton is~$H$.%
\footnote{In a simple graph, a trivial block is always a bridge in $G$. In multigraphs,
$H$ may also be a set of parallel edges between a vertex pair $x,y$, whose removal would disconnect $x,y$. 
In the latter case, it is trivial that
the precise order of the edges is irrelevant, and we could represent $H$ 
via a single edge $e_H$ and its multiplicity $|E(H)|$ instead.
}
Also, let the skeleton $S_\nu$ of a C-node $\nu$ simply be the corresponding cut vertex.\defend
\end{defn}
Clearly, the linear-sized con-tree $\scri C(G)$ can be obtained from $G$ in linear time.

Furthermore, we will sometimes treat this two-level con-tree as follows:
\begin{defn}[Flattening a con-tree, decomposition graph]
\label{def:flat-con-tree}
In the setting of Definition~\ref{def:con-tree},
let a \emph{decomposition graph} $\scri D(G)$
of $G$ be constructed from the union of all the sSPR trees $\scri T(H)$ 
over the blocks $H$ of $G$ and of all the C-nodes of $\scri B(G)$, as follows:
\begin{enumerate}[i)]
\item
Observe that for each block $H$ in $G$ holding a cut vertex $x$ of $G$, 
there is one or possibly multiple nodes of $\scri T(H)$ whose skeletons contain
$x$, and not all of them are P-nodes. 
Call these nodes of $\scri T(H)$ whose skeletons contain
$x$ the {\em mates of $x$} from $\scri T(H)$.
\item
In $\scri D(G)$, for every C-node $\mu$ holding a cut vertex $x$ of $G$
and for every block $H$ of $G$ containing~$x$,
make $\mu$ {\em adjacent} to all the mates of $x$ from $\scri T(H)$ 
that are not P-nodes.
\defend
\end{enumerate}
\end{defn}
This ``flat'' view of a decomposition graph will later be implicitly used,
for instance, when defining a con-path (Definition~\ref{def:Con-path}).
The intentional exclusion of P-nodes in (ii) is of technical nature and 
will be justified later in Claim~\ref{clm:Pnodes}.

\smallskip
To summarize, a con-tree (as well as the related decomposition graph)
provides us with a natural and easy way to encode all
embeddings of a given planar graph. An S-node skeleton has a unique embedding.
For an R-node, we can choose between the unique embedding of 
its skeleton and the mirror image.
For a P-node, we can choose an arbitrary cyclic permutation of the skeleton
edges.
For a C-node, a precise description of the choice is more tricky,
but it roughly means we can choose which face of each block
(assuming the blocks are already embedded) holds which other block.
On the other hand, given a particular plane embedding of a graph $G$,
it is trivial to deduce the embeddings of all the skeletons within the
con-tree $\scri C(G)$---we shall formalize this observation in
Definition~\ref{def:embedding-spec} and Claim~\ref{clm:embedding-spec}.

\subsection{Con-chains and con-paths}

In this paper we refer to the optimal single-edge insertion algorithm
by Gutwenger et al.~\cite{GMW05}.
Consider (cf.~Definition~\ref{def:MEI} for $k=1$)
a connected planar graph $G$ and let $v_1,v_2$ 
be the vertices we want to connect by a new edge.
In a nutshell (see Section~\ref{sec:singleIns} for details),
the algorithm of~\cite{GMW05}
shows that an optimal plane embedding (of $G$) for inserting $v_1v_2$ into $G$
depends only on the con-tree nodes ``between'' some node containing $v_1$
and one containing $v_2$.
In this regard, we bring the following specialized definitions.

\begin{defn}[Con-chain]
\label{def:con-chain}
Consider a connected planar graph $G$, and its con-tree $\scri C(G)$.
Let $v_1,v_2\in V(G)$.
The {\em con-chain} $\scri Q_{v_1v_2}$ of the pair $\{v_1,v_2\}$ in $\scri C(G)$ can be uniquely defined as consisting of the following:
\begin{enumerate}[(i)]
\item the unique shortest path $Q$ in $\scri B(G)$ between a B-node
$\beta_1$ whose block contains $v_1$ and a B-node $\beta_2$ whose block contains $v_2$;
\item for each B-node $\beta$ along $Q$,
two {\em border vertices} $w_1^\beta,w_2^\beta$: for $i=1,2$, if $\beta=\beta_i$ let
$w_i^\beta=v_i$; otherwise let $w_i^\beta$ be the cut vertex in $G$ corresponding to the (unique) C-node neighbor 
of $\beta$ closest to $\beta_i$;%
\footnote{Note that $w_2^\beta=w_1^{\beta'}$ if $\beta,\beta'$ are
consecutive B-nodes on $Q$ in this order.} and
\item\label{it:con-chain-iii} again for each $\beta$ on $Q$,
the node sequence $Q_\beta$ of the unique shortest path in $\scri T(H_{\beta})$ 
connecting nodes of $\scri T(H_{\beta})$ whose skeletons contain
the border vertices $w_1^\beta$ and $w_2^\beta$ in this order. 
\end{enumerate}
In \ref{it:con-chain-iii}, uniqueness of the path generally follows from the fact that
$\scri T(H_{\beta})$ is a tree.
For the possible degenerate case that $Q_\beta$ consists of a single node
(whose skeleton holds both $w_1^\beta,w_2^\beta$) but is not unique,
we choose the ``first'' such node, according to some arbitrary
but fixed order of the decomposition nodes.
(In fact, this special case is showcased in Figure~\ref{fig:exampledecomp}.)\defend
\end{defn}

\begin{defn}[Con-path]
\label{def:Con-path}
In the setting of Definition~\ref{def:con-chain}, 
a {\em con-path} \linebreak$\scri P_{v_1,v_2}:=\scri P(\scri Q_{v_1,v_2})$
of\/ $\{v_1,v_2\}$ in $\scri C(G)$ is
the sequence of decomposition nodes (of type S, P, R, C, and D) 
resulting from $Q$ (which contains B- and C-nodes), by replacing 
each B-node $\beta$ by its corresponding sequence $Q_\beta$. 
\defend
\end{defn}

Obviously, no con-path may start or end with a C-node. Moreover:
\begin{claim}\label{clm:Pnodes}
A con-chain piece $Q_\beta$ (cf.\ Def.~\ref{def:con-chain}(iii)) can neither start nor end with a P-node. 
Therefore, a P-node can occur in a con-path only between two S-nodes,
and never at the end nor adjacent to a C-node.
\qed\end{claim}
The term \emph{con-path} is justified by the fact that this node sequence is a
path 
in the decomposition graph $\scri D(G)$ from Definition~\ref{def:flat-con-tree}.
We may consider a con-path implicitly oriented, when suitable.

Even though a decomposition graph $\scri D(G)$ is not a tree,
our con-paths in $\scri D(G)$ ``behave'' similarly as paths do in a tree:

\begin{claim}\label{clm:con-intersection}
Consider two con-paths $\scri P_1$, $\scri P_2$ of pairs $\{v_1,w_1\}$ and
$\{v_2,w_2\}$ in the decomposition graph $\scri D(G)$ of a graph $G$.
Then $\scri P_1\cap\scri P_2$, if non-empty, forms a consecutive subsequence
in each of $\scri P_1,\scri P_2$.
\end{claim}

\begin{proof}
In this proof we view $\scri P_1,\scri P_2$ as paths in the graph $\scri D(G)$.
Assuming a contradiction, there exist distinct subpaths 
$\scri P_i'\subseteq \scri P_i$, $i=1,2$, such that $\scri P_1', \scri P_2'$ 
share common ends $\alpha,\beta\in V(\scri D)$ but are internally disjoint
otherwise.
Since C-nodes are cut nodes in $\scri D(G)$ by Definition~\ref{def:flat-con-tree},
$\scri P_1'\cup\scri P_2'$ contains no C-nodes except possibly $\alpha$
and/or $\beta$.
If neither of $\alpha,\beta$ is a C-node, then $\scri P_1', \scri P_2'$
are contained in the same sSPR-tree of $\scri C(G)$ 
and so $\scri P_1'=\scri P_2'$, a contradiction.
Similarly, if both $\alpha,\beta$ are C-nodes then by uniqueness claimed in
Definition~\ref{def:con-chain}.\ref{it:con-chain-iii}, it is
$\scri P_1'=\scri P_2'$.

It remains to consider that, up to symmetry,
$\alpha$ is a C-node and $\beta$ belongs to an sSPR-tree $\scri T(H)$ where
$H\subseteq G$ is a block incident with $a\in V(G)$, the cut vertex of $\alpha$.
Let $\alpha_i$ be the neighbor of $\alpha$ on $\scri P_i'$, $i=1,2$.
Then $\beta$ lies on the unique $\alpha_1$-$\alpha_2$ path in $\scri T(H)$,
and since both the skeletons $S_{\alpha_1},S_{\alpha_2}$ contain $a$,
it is $a\in V(S_\beta)$.
Unless $\beta$ is a P-node, $\scri P_1'=\scri P_2'$ is thus formed by a
single edge $\alpha\beta$.
In the case of $\beta$ being a P-node, cf.~Claim~\ref{clm:Pnodes},
$\beta$ is not an end of either $\scri P_1,\scri P_2$.
Therefore, if $\beta_i\in V(\scri P_i)\setminus V(\scri P_i')$ is the
(other) neighbor of $\beta$ on $\scri P_i$, $i=1,2$, we have
$a\in V(S_{\beta_i})$ and by minimality in
Definition~\ref{def:con-chain}.\ref{it:con-chain-iii} it is 
$\beta\not\in V(\scri P_i)$, a contradiction again.
\qed\end{proof}

\subsection{Towards ``embedding preferences''}
\label{sec:towardse}

On a high level, the starting point of our algorithm is to solve the single 
edge insertion problem for each edge $\{v_1,v_2\}\in F$ into $G$ independently.
In this way we obtain so called \emph{embedding preferences}
(Definition~\ref{def:embedding-pref-wrt}) for each node $\nu\in\scri P_{v_1,v_2}$. 
Equipped with all these preferences for all edges of $F$, we have to decide on an 
embedding $G_0$ of $G$ which, essentially, ``honors'' as many of these 
preferences as possible. It should be understood that the preferences for 
distinct edges of $F$ may naturally conflict with one another. We will ensure 
that we pick an embedding $G_0$ which guarantees that any $\{v_1,v_2\}\in F$ can be 
inserted into $G_0$ (as a new edge $v_1v_2$) without ``too many'' additional crossings, 
compared to its individual optimum $\ins(G,v_1v_2)$.

The exact scheme for our embedding preferences is a critical part of the
proof of our algorithm, 
and in order to explain our complete definitions (Section~\ref{sub:embedcontree}), 
it seems worthwhile to discuss the situation and drawbacks of possible 
alternatives beforehand. 
We start with a summary of the algorithm \cite{GMW05} to optimally insert a single edge.
Our description is very informal since we never use the details of this
algorithm in our paper, but the summary will help us to explain the
(somehow unexpected) deep technical problems connected with our extension to
multiple edges.

\paragraph{Single edge insertion algorithm \cite{GMW05}.}
We can interpret the algorithmic steps of~\cite{GMW05} for $\ins(G,v_1v_2)$ as
what we call {simple embedding preferences}. 
In our terminology of Definition~\ref{def:con-chain}, the 
insertion algorithm for the edge $v_1v_2$ first considers individually
each B-node $\beta$ of $\scri Q_{v_1,v_2}$. 
Let $Q_\beta=(\nu_1,\nu_2,\ldots,\nu_r)$ 
be ordered such that $\nu_1$ contains $w_1^\beta$  and $\nu_r$ contains $w_2^\beta$. 
For $\nu_i\in Q_\beta$, one independently computes an embedding and 
a local insertion path as follows.
 
The source (target) of the local insertion path is either the vertex 
$w_1^\beta$ ($w_2^\beta$), if $i=1$ ($i=r$), or otherwise $e_{\nu_{i-1}}$ 
($e_{\nu_{i+1}}$) in $S_{\nu_i}$\,---the virtual edge(s) corresponding to the
predecessor and the successor of $\nu_i$, respectively.
The insertion path is computed in the dual of the skeleton $S_{\nu_i}$:
For S-nodes, the path is simply one of the two faces, and no crossings arise. 
Similarly, no crossings arise for P-nodes; 
one chooses a skeleton embedding in which the source and the target lie on a
common face. This face then constitutes the insertion path. 
In case of an R-node, the embedding of $S_{\nu_i}$ is fixed and 
one chooses any shortest path in its dual as the insertion path. 
Crossins may, unavoidably, arise in this case.

In each of the cases, if the source or the target is a virtual edge, 
one also notes whether the insertion path attaches to it from the left or
the right side (left/right with respect to an arbitrarily fixed but consistent
direction of all the virtual edges).
This solution and note together constitute the {\em simple embedding preference} 
for each node of~$Q_\beta$ in the con-tree of~$G$.

All the simple embedding preferences along $Q_\beta$ can be simultaneously
satisfied, by stepwise ``gluing together'' the individual subembeddings
along twin pairs of virtual edges:
this is trivial when the local insertion paths of adjacent nodes 
attach to the twin pair from different sides, and one
flips the next skeleton embedding before gluing in the other case.
For other nodes without preferences, one picks any embedding,
as the insertion solution is independent of this decision.

Having embedded the blocks of~$G$, it is now easy to deal with C-nodes. 
For any C-node $\gamma$ of $\scri Q_{v_1,v_2}$ (corresponding to a cut vertex
$x\in V(G)$), let $H,H'\subseteq G$ be the incident blocks.
The corresponding ``loose ends'' of the previously computed insertion paths
in $H$ and $H'$ attach to $x$ from certain faces $\varphi,\varphi'$
of embedded $H,H'$, respectively.
In an embedding of the union $H\cup H'$ one identifies $\varphi$ with
$\varphi'$, such that the insertion paths are joined together without     
additional crossings.
All remaining blocks can be joined arbitrarily as the insertion solution 
is independent of these decisions. 

\paragraph{Properties of good preference schemes.}
In our case of multiple edges, the above description has inherent problems, 
some rather obvious, some more subtle.
First, simple prefereces require a stepwise resolution
(i.e., having to decide all the blocks, before being able to properly define C-node
preferences), which does not allow for an easy way of combining multiple
insertion paths. 

Second, simple preferences are capable of handling only a quite restricted
class of embeddings.
In our case, this would inevitably lead to a situation in which two
distinct insertion paths set different preferences for the nodes they follow
together, even though there exists an equally optimal solution for one of
the paths that nearly matches the preferences of the other path.
A very simple example of this is when the first path decides for an
embedding $G_0$ while the second one decides for the mirror image of~$G_0$.
Note, however, that there is no big problem with routing through rigid
skeletons of R-nodes, as one can always define a unique canonical order on
all the dual paths there, and a unique default choice between the skeleton
and its mirror image (flip).

The aforemention problem gets tougher when considering S- and P-nodes
(and also C-nodes in greater generality).
Consider that an embedding of a P-node $\nu$ skeleton is specified by a clockwise
order of its edges (yes, it is actually enough to specify a pair of consecutive edges).
However, another specification at $\nu$ may reverse the edge order of the $\nu$
skeleton and simultaneously flip both the adjacent skeleton embeddings.
This results in, essentially, the same embedding of $G$, and has to be
captured as ``the same'' by our new preference scheme.
But then we would mix the concepts of specifying subembeddings 
(e.g., the edge order at P-nodes) and of handling the gluing operation (flips).
In this perspective, our extension from SPR- to sSPR-trees can be seen as 
a first step to decouple flipping decisions from ordering decisions within
P-nodes---it allows us to encode, at a mandatory S-node, whether the adjacent
skeletons are flipped ``the same way'' or ``the other way'';
with special labels \lnonswitching/\lswitching.

Though, the problems are not over yet.
One would also, naturally, like to define the preferences so that it is \emph{trivial} 
to validate if some given embedding satisfies the preferences or not. 
In other words, one would like that every bit of information 
stored as a preference has a directly observable counterpart in the final
embedding~$G_0$.
While there are ways to rigorously define a suitable preference scheme achieving
this goal,%
\footnote{For example, it is the preference scheme originally described in the 
conference version of this research \cite{CHicalp}.}
the solution is not satisfactory for the following reason.
With any directly observable preference scheme, it seems inevitable that
validation of the embedding preference of one node requires us to look at
the embedding preferences of some of its neighbors (sometimes up to distance two).
Not surprisingly, this would bring big complications for the proofs,
resulting in long (and difficult to verify) case-checking arguments.

In our new preference scheme, we hence trade the (part of) direct
observability property for a new property of {\em strong locality}---demanding
that validation of the embedding preference of one node is possible without
looking at the preferences of any other nodes.
For this purpose we shall introduce a new kind of ``unobservable''
information into an embedding---a so called spin (of virtual edges),
which helps to achieve the goal of strong locality by providing us with kind
of a ``communication channel'' between neighboring preferences.
All of this is the subject of the following rather technical
Section~\ref{sub:embedcontree}.

\subsection{Embedding preferences in a Con-tree}
\label{sub:embedcontree}

In this section we give the formal definitions establishing our scheme
of embedding preferences for con-chains that achieves the goal of strong
locality, as informally explained in Section~\ref{sec:towardse}.

We start with formally specifying what an embedding is
(Definition~\ref{def:embedding-spec}) and what the partially unobservable
information we add to it is (Definition~\ref{def:spins} and Claim~\ref{clm:con-embedding-spin}).
Then, we define the exact pieces of information we are going to store at a
con-chain node as its embedding preference
(Definition~\ref{def:embedding-pref-wrt}), and we couple this definition
with a specification (respecting strong locality) of what it means that a
certain collection of embedding preferences is honored by an embedding
of~$G$ (Definition~\ref{def:embedding-honor}).

For reference in the following definitions, we need to define certain ``defaults''
for embeddings of a graph~$G$.
Let $G_d$ be an arbitrary but {\em fixed} embedding of $G$ in the plane
(i.e., including the specification of the unbounded face).
From $G_d$ we derive (a) the \emph{default embedding} for each R-node skeleton
(distinguishable from its mirror), and (b) the \emph{default face} 
for each S-node skeleton (distinguishable from its other face) 
by picking the bounded face of this skeleton in $G_d$.
Note that the default face is well-defined even for an S-node skeleton forming a
$2$-cycle since the skeleton edges are labeled.
Specially, for each D-node skeleton (a bunch of parallel non-virtual edges), 
we observe that it is always possible to draw all its edges close together
such that the rest of $G$ is embedded in one face of it and that no
crossings with the inserted edges arise.
This chosen face, in any embedding of $G$, will be called the \emph{default face}
of a D-node skeleton (while disregarding the names and order of the
skeleton edges).

\begin{defn}[Embedding specification]
\label{def:embedding-spec}
 Consider a connected planar graph $G$, an embedding $G_1$ of $G$, and 
 the con-tree $\scri C:=\scri C(G)$.
 % and its decomposition graph $\scri D:=\scri D(G)$. 
 Let an {\em embedding specification} $\scri E(G_1)$
 be a collection of information specified as follows:
 \begin{enumerate}[i)]
  \item For each R-node $\nu$ of $\scri C$, we specify one of the labels \lflipped or \lnonflipped;
 the information is \lnonflipped exactly when the embedding of $S_\nu$
 induced by $G_1$ is the default one.
  \item For each P-node $\nu$ of $\scri C$, we specify (an arbitrary) one of the two vertices of
 $V(S_\nu)$ and the cyclic clockwise order of the edges of $S_\nu$ around it
 in the embedding of $S_\nu$ induced by $G_1$.
  \item For each C-node $\nu$ of $\scri C$ and each pair $H_1,H_2\subseteq G_1$ of incident blocks,
 we specify the face of $H_j$ that contains $H_{3-j}$ in the embedding
 $G_1$, for~$j=1,2$.
\end{enumerate}
\end{defn}
An embedding $G_1$ uniquely determines
the embedding specification~$\scri E(G_1)$ by definition.
Conversely, it immediately follows that:
\begin{claim}\label{clm:embedding-spec}
If $G_1,G_2$ are two embeddings of a connected planar graph $G$ such that
$\scri E(G_1)=\scri E(G_2)$, then $G_1$ and $G_2$ are equivalent (and also not mirrored
images of each other).
\qed\end{claim}

While this embedding \emph{specification} of Definition~\ref{def:embedding-spec}
cannot be used directly as a basis for our embedding \emph{preference} scheme,
it serves as a blue print of what our preferences need to model.

Recall our discussion from Section~\ref{sec:towardse} about flipping
skeleton embeddings during the gluing operation of adjacent skeletons.
For the intended purpose of decoupling the flipping decision of a glue
operation of Definition~\ref{def:SPR} from the embedding decision for
skeletons, we introduce the concept of \emph{spins} in an embedding.
Interestingly, in general, spin values are not fully determined by a particular embedding.
A spin can also be (informally) seen as a 1-bit communication channel transferring certain
information between nonadjacent con-tree nodes, 
which will allow us to establish the desired strong locality of embedding preferences.

\begin{defn}[Enriched embedding, spins]
\label{def:spins}
 Consider a connected planar graph $G$, its embedding $G_1$ and the con-tree $\scri C(G)$.
 A \emph{CS-pair} is a tuple $(c,\nu)$ where $c$ is a cut vertex in $G$ and $\nu$ is 
an S-node of $\scri C(G)$ whose skeleton contains~$c$. 
 In an \emph{enriched} embedding $\hat{G}_1$ of $G_1$, we charge 
each virtual edge and each CS-pair of the con-tree $\scri C(G)$
 with either a \emph{positive} or a \emph{negative spin}, subject to the following properties:
 \begin{itemize}
 \item[] \begin{enumerate}[(E1)]
 \item A virtual edge and its twin will always be charged identically.
 \item For an R-node $\varrho$,
 if the embedding of the skeleton $S_\varrho$ is the default one
 (i.e., $\varrho$ is labeled \lnonflipped in $\scri E(G_1)$),
 then all the virtual edges in $S_\varrho$ are charged with a positive spin;
 otherwise, all those virtual edges are charged with a negative spin.\defend
 \end{enumerate}
 \end{itemize}
\end{defn}
The role of CS-pairs $(c,\nu)$ deserves a further informal explanation.
First, note that $c$ in such a pair is a mate of $\nu$ in the sense of
Definition~\ref{def:flat-con-tree}.
Second, a CS-pair will be used conceptually similarly to a pair
of twin virtual edges of an sSPR tree---while twin pairs of virtual edges
are used to glue pieces of an sSPR tree together, CS-pairs will be used to
specify the precise way two adjacent blocks are glued together at a cut vertex.
Though, we will need this additional information only in the case
when $\nu$ is an S-node (and not for R-nodes for which we may explicitly
refer to a certain face).

Since R-nodes are never directly adjacent, there cannot arise any 
conflicting spin values from Definition~\ref{def:spins}. %at any virtual edge.
Most importantly, all spin values not predeterminded by properties E1 and E2 are only loosely correlated 
with the embedding $G_1$, and they will \emph{not} be part of the embedding preference.
Their role is to function as a mediator between our embedding preferences and an actual embedding.
In other words, we will only ask whether \emph{there exist}
spin values such that an embedding satisfies our preferences, and not what
the spin values are.

\medskip
The following two entangled definitions achieve all our goals concerning embedding
preferences in a con-tree and their interpretation.

\begin{defn}[Node embedding preferences for a con-chain]~\\
\label{def:embedding-pref-wrt}%
Consider a connected planar graph $G$, its con-tree $\scri C(G)$, and the decomposition graph 
$\scri D:=\scri D(G)$ (Definition~\ref{def:flat-con-tree}).
An {\em embedding preference $\pi_\nu$ of a node} $\nu\in V(\scri D)$ is defined
(only) if $\nu$ is an S-, P-, or C-node.
One piece of information that $\pi_\nu$ stores
is a pair of distinct nodes $\mu_1,\mu_2\in V(\scri D)$,
called the {\em $\pi_\nu$-peers of~$\nu$}, such that $\mu_1,\mu_2$ are neighbors of
$\nu$ in~$\scri D$.
Furthermore:
 \begin{itemize}
 \item[] \begin{enumerate}[\bf (P1)]
 \item\label{def:it:prefP}
	 If $\nu$ is a P-node, we only store these peers $\mu_1,\mu_2$ 
	(which, in this case, are S-nodes by definition) in $\pi_\nu$.
 \item\label{def:it:prefS}
	If $\nu$ is an S-node, we additionally store in $\pi_\nu$ one of the
	labels \lswitching or \lnonswitching.
	However, if neither $\mu_1$ nor $\mu_2$ is an R-node, the
	stored label must always be \lnonswitching.
 \item\label{def:it:prefC}
	If $\nu$ is an C-node, then the stored $\pi_\nu$-peers $\mu_1,\mu_2$
        are R-, D- or S-nodes by definition.
	For $i=1,2$,
	if the peer $\mu_i$ is an R-node, then we additionally store in
	$\pi_\nu$ a label specifying a face in the skeleton~$S_{\mu_i}$.
 \end{enumerate}
 \end{itemize}
Consider a non-adjacent vertex pair $v_1,v_2\in V(G)$ and its con-chain
$\scri Q_{v_1,v_2}$ in $\scri C(G)$.
Let $\scri P:=\scri P(\scri Q_{v_1,v_2})\subseteq\scri D$ be the
corresponding con-path.
An {\em embedding preference $\Pi_{v_1v_2}$ of $\scri Q_{v_1,v_2}$}
(equivalently, a {\em preference of $\scri P$})
is a collection of embedding preferences $\pi_\nu$ over those
internal nodes $\nu$ of $\scri P$ that are neither R- nor D-nodes,
subject to the following restriction:
 \begin{itemize}
 \item[] \begin{enumerate}[\bf (P4)]
 \item\label{def:it:prefN}
	For every individual preference $\pi_\nu$ in $\Pi_{v_1v_2}$,
	the two $\pi_\nu$-peers of $\nu$ must be the two neighbors of $\nu$ on~$\scri P$.
\defend
 \end{enumerate}
\end{itemize}
\end{defn}
For consistency, we say that a node $\nu$ whose embedding preference is not defined
in Definition~\ref{def:embedding-pref-wrt} ($\nu$ might be an R- or D-node),
has a {\em void} embedding preference.
Then,
we can rigorously speak about embedding preferences
of {\em all} the internal nodes of a con-path $\scri P$, 
or of all nodes in an arbitrary subset of $V(\scri D)$.

\begin{defn}[Honoring an embedding preference]%
\label{def:embedding-honor}~\\
Consider a connected planar graph $G$ and any embedding $G_1$ of $G$, 
specified via its embedding specification $\scri E:=\scri E(G_1)$. 
Let $U$ be a subset of nodes of $\scri C(G)$ and
$\Pi_U$ be a collection of (arbitrary) %, and possibly void) 
node embedding preferences $\pi_\nu$ over $\nu\in U$.
We say that $G_1$ \emph{honors the embedding preferences $\Pi_U$} if
there \emph{exists} an enriched embedding $\hat{G}_1$ of $G_1$ such that,
for every $\nu\in U$ where $\pi_\nu$ is non-void,
$\hat{G}_1$ is {good for $\pi_\nu$}.
That is, $\hat{G}_1$ is {\em good for $\pi_\nu$}
(where $\nu$ falls into one of the cases (P1)--(P3) of
Definition~\ref{def:embedding-pref-wrt}) if the appropriate one of the following
cases holds:
 \begin{description}\parskip3pt
\item[case \ref{def:it:prefP}:] $\nu$ is a P-node.
Let $\mu_1,\mu_2$ be the $\pi_\nu$-peers (both being S-nodes) of~$\nu$, and let $e_{\mu_1}$,
$e_{\mu_2}$ denote the virtual edges in $S_\nu$ corresponding to $\mu_1$,
$\mu_2$, respectively.
The cyclic order of the edges in $S_\nu$ specified in $\scri E$ is such
that $e_{\mu_1}$ and $e_{\mu_2}$ occur as neighbors---they form a 
common face $\varphi$ in $S_\nu$. 
Moreover, for each $i = 1,2$, the face $\varphi$ corresponds
to the default face of $S_{\mu_i}$ if $e_{\mu_i}$ has the 
positive spin in $\hat{G}_1$, and $\varphi$ corresponds to the other face 
of $S_{\mu_i}$ if the spin is negative.

\item[case \ref{def:it:prefS}:] $\nu$ is an S-node.
There are two spin values associated with $\pi_\nu$ via $\hat{G}_1$: for $i=1,2$, the $i$-th
spin value is the one charged to the virtual edge $e_{\mu_i}$ if the
$\pi_\nu$-peer $\mu_i$ is an R- or P-node,
and otherwise the $i$-th spin value is the one charged to the CS-pair formed by
$\nu$ and the cut vertex of the C-node~$\mu_i$.
If the label in $\pi_\nu$ is \lswitching then exactly one of these spin values 
of $\hat{G}_1$ is positive and the other one negative, while if the label is
\lnonswitching then both these spin values have to be the same.%
\footnote{Note that (some of) the spin values in $\hat{G}_1$ are correlated
with the specification $\scri E$ according to Definition~\ref{def:spins},
and so $\scri E$ is implicitly considered in the case 
{\let\bf\relax\ref{def:it:prefS}}, too.}

\item[case \ref{def:it:prefC}:] $\nu$ is a C-node.
Let $H_1,H_2\subseteq G$ be the blocks the $\pi_\nu$-peers $\mu_1,\mu_2$ of $\nu$ belong to, 
and let $c$ be the cut vertex corresponding to $\nu$. 
There are two skeleton faces, $\varphi_1$ of $S_{\mu_1}$ and $\varphi_2$ of $S_{\mu_2}$,
associated with $\pi_\nu$ via $\hat{G}_1$ as follows.
Let $i\in\{1,2\}$.
If $\mu_i$ is an R-node then $\varphi_i$ is explicitly given as a label in $\pi_\nu$.
If $\mu_i$ is a D-node then $\varphi_i$ is the default face of $S_{\mu_i}$.
If $\mu_i$ is an S-node, then $\varphi_i$ is the default face of $S_{\mu_i}$
in the case that the CS-pair $(c,\mu_i)$ is charged with a positive spin,
and otherwise (negative spin) $\varphi_i$ is the other face of $S_{\mu_i}$.
The two faces specified in $\scri E$ for $H_1,H_2$ at $\nu$ must correspond to $\varphi_1$ and 
$\varphi_2$ (disregarding their orientation), respectively.
\defend
 \end{description}
\end{defn}

Void embedding preferences from $\Pi_U$, and possible preferences
of other nodes not in $U$, do not matter when deciding whether an embedding $G_1$
honors $\Pi_U$ as above.
We also call any enriched embedding $\hat{G}_1$ of $G_1$ that satisfies all the
requirements of Definition~\ref{def:embedding-honor} a {\em good enriched
embedding for $\Pi_U$}.

\begin{claim}
\label{clm:con-embedding-spin}
Assume that a plane embedding $G_1$ of $G$ honors embedding preferences
$\Pi:=\Pi_{v_1v_2}$ of a con-chain $\scri Q_{v_1,v_2}$.
Then, the spin values charged to the virtual edges and the CS-pairs 
along $\scri Q_{v_1,v_2}$ are consistent over all possible good enriched embeddings
for $\Pi$.
These spin values can be derived directly from the embedding specification
$\scri E(G_1)$ (w.r.t.\ the implicit default embedding $G_d$ of~$G$).
\end{claim}
\begin{proof}
Since $G_1$ honors $\Pi$, 
there exists at least one feasible spin assignment by definition.
We are going to show that the spin values are uniquely determined 
along the con-path~$\scri P:=\scri P(\scri Q_{v_1,v_2})$.

Consider any edge $\nu\mu\in E(\scri P)$.
If one of $\nu,\mu$ is an R-node (and the other one is not a C-node),
then the two virtual edges associated with $\nu\mu$ in $\scri C(G)$
are charged a spin value that is uniquely determined by
Definition~\ref{def:spins}.

Assume that, say, $\nu$ is a C-node of a cut vertex~$c$.
Then a spin is charged to the CS-pair $(c,\mu)$ only if $\mu$ is an S-node.
We determine the spin value of $(c,\mu)$ by comparing
the face specified in $\scri E(G_1)$ for $\nu$ and the block of $S_\mu$ with
the default face of $S_\mu$, as claimed in 
Definition~\ref{def:embedding-honor}, case {\let\bf\relax\ref{def:it:prefC}}.

The remaining case to consider is, up to symmetry, that $\nu$ is a P-node
and $\mu$ an S-node.
The spin value of the twin virtual edges $e_\mu,e_\nu$ (in $S_\nu,S_\mu$,
respectively) is again determined by comparing the embedding of
the P-skeleton $S_\nu$ in $\scri E(G_1)$ (precisely, its face
specified by the preference $\pi_\nu$) to the default face of $S_\mu$,
as claimed in 
Definition~\ref{def:embedding-honor}, case {\let\bf\relax\ref{def:it:prefP}}.
\qed\end{proof}

The consequence of the above claim is very interesting. On the one hand, it tells
us that, effectively, we do not need to care too much about enriched embeddings from now on,
except for in ``low-level'' proofs. On the other hand, however, 
it is \emph{not} straightforwardly possible to remove the spins
from the definition of embedding preferences altogether.
A key problem occurs with an S-node $\nu$ and its virtual edge $e_\mu$
such that an embedding preference of the neighbor $\mu$ does not refer to
$\nu$ (but to other nodes): then, the spin value of $e_\mu$ is undetermined
and represents a ``communication channel'' between the embedding
preferences of adjacent nodes.
Abandoning this vital ``communication link'' would require us,
e.g., for a P- or C-node in Definition~\ref{def:embedding-honor}, to refer
also to the embedding preferences of the neighbors and thus
would violate our goal of strong locality
(with some nasty consequences in the later proofs).

\subsection{Single edge insertion and embedding preferences}
\label{sec:singleIns}

Note that it is not a priori clear, given the additional conditions in
Definitions \ref{def:embedding-pref-wrt},\,\ref{def:embedding-honor} and the
elusive nature of spins,
that our treatment of embedding preferences is capable of rigorously
describing every optimal solution to the single edge insertion problem.
It is the task of coming Lemma~\ref{lem:correctpref} to prove this
important and nontrivial finding.
(While perhaps looking similar to Claim~\ref{clm:con-embedding-spin}, this
task is in fact very different from the former one.)

\begin{lemma}
\label{lem:correctpref}
Let $G$ be a connected planar graph and $v_1,v_2\in V(G)$.
Assume that $G_0$ is a plane embedding of $G$
such that a new edge $f=v_1v_2$ can be drawn into $G_0$ with $\ins(G,f)$ crossings.
Then there exists an embedding preference $\Pi_{v_1v_2}$ 
of the con-chain $\scri Q_{v_1,v_2}$
such that $G_0$ honors $\Pi_{v_1v_2}$.
\end{lemma}

\begin{proof}
Let $\scri P:=\scri P(\scri Q_{v_1,v_2})$ be the corresponding con-path,
and $\scri E=\scri E(G_0)$ be the embedding specification of~$G_0$.
There are two tasks in this proof:
\begin{enumerate}[a)]
\item to deduce individual embedding preferences $\pi_\nu$
(forming $\Pi_{v_1v_2}$)
for all internal nodes of $\scri P$, and
\item to specify an enriched embedding $\hat{G}_0$ such that all the
conditions of Definition~\ref{def:embedding-honor} are satisfied for
$\Pi_{v_1v_2}$ via $\hat{G}_0$.
\end{enumerate} 

As for a), we easily determine all information except the switching
attributes of the internal S-nodes on $\scri P$ from
Definitions~\ref{def:embedding-pref-wrt},\,\ref{def:embedding-honor}:
the $\pi_\nu$-peers are the neighbors of $\nu$ on $\scri P$
and, in the case of a C-node $\nu$ with an R-neighbor, the face label(s)
follows Definition~\ref{def:embedding-honor}, case {\let\bf\relax\ref{def:it:prefC}}.

Assuming for a moment that, for each internal P-node $\nu$ of $\scri P$,
the two virtual edges of the $\pi_\nu$-peers form a face in $\scri E$,
in b) we can determine spin values along $\scri P$ exactly as in the
proof of Claim~\ref{clm:con-embedding-spin}.
Remaining spin values can be charged arbitrarily, giving an enriched
embedding $\hat{G}_0$ of $G_0$.
Returning back to a), the switching attribute of each $\pi_\nu$ where $\nu$ is
an S-node now follows from charged spin values and 
Definition~\ref{def:embedding-honor}, case {\let\bf\relax\ref{def:it:prefS}}.

It remains to verify three facts;
our assumption about P-nodes of $\scri P$, and fulfillment of
Definition~\ref{def:embedding-pref-wrt}\,{\let\bf\relax\ref{def:it:prefS}}
and of Definition~\ref{def:embedding-honor}.
This can be done conveniently along the cases of the latter definition.
Let $\nu$ be an internal node~of~$\scri P$
and $\mu_1,\mu_2$ be the two $\pi_\nu$-peers of $\nu$ (its neighbors
on~$\scri P$).

\begin{description}%\parskip3pt
\item[case \ref{def:it:prefP}:] $\nu$ is a P-node.
For $i=1,2$, let $H_i\subseteq G$
be the subgraph obtained by recursively gluing
skeletons to the virtual edges of $S_{\mu_i}$ except to $e_{\nu}$.
Up to symmetry between the indices $i=1,2$, either $v_i\in V(H_i)$
or $v_i$ is separated from $V(S_{\mu_i})\cup\{v_{3-i}\}$ 
by a cut vertex $c$ of $G$ such that $c\in V(H_i)$.
Notice that $V(H_1)\cap V(H_2)=\{p_1,p_2\}$ is a split pair in $G$ 
formed by $V(S_\nu)=\{p_1,p_2\}$.
Obviously, the edge $f$ in optimal $G_0+f$ does not cross 
any component represented by a virtual edge of $S_\nu$, other than possibly $H_1,H_2$.
Therefore, the virtual edges $e_{\mu_1},e_{\mu_2}$ indeed form a face
in the embedding specification~$\scri E$ of~$G_0$.

\item[case \ref{def:it:prefS}:] $\nu$ is an S-node.
The switching label of $\pi_\nu$ has already been chosen to satisfy
Definition~\ref{def:embedding-honor} via $\hat{G}_0$.
As in the previous case, the edge $f$ in optimal $G_0+f$ obviously does not cross    
any of the components represented by the virtual edges of $S_\nu$ other than
possibly $e_{\mu_1},e_{\mu_2}$.
Hence, unless one of $\mu_1,\mu_2$ is an R-node (in which case
$e_{\mu_1}$ or $e_{\mu_2}$ has been charged a spin by Definition~\ref{def:spins}),
both the spin values charged for $\mu_1$ and $\mu_2$
(each may be of a virtual edge $e_{\mu_i}$ or of a CS-pair~$(x_i,\nu)$) 
refer to the same face of $S_\nu$ and the label of $\pi_\nu$ is \lnonswitching, 
in agreement with Definition~\ref{def:embedding-pref-wrt}.

\item[case \ref{def:it:prefC}:] $\nu$ is a C-node.
In this case there is nothing more to discuss since $\pi_\nu$ and
$\hat{G}_0$ have already been chosen to satisfy Definition~\ref{def:embedding-honor}.
\smallskip\qed
\end{description}
\end{proof}

In the algorithmic context, 
the key is to find such an embedding preference $\Pi_{v_1v_2}$ as in
Lemma~\ref{lem:correctpref} efficiently.
Using Lemma~\ref{lem:correctpref}, we translate the main result of \cite{GMW05} 
into our slightly different setting as follows:

\begin{theorem}[Gutwenger et al.~\cite{GMW05}, reformulated]
\label{thm:singleedge}
Let $G$ be a connected planar graph and $v_1,v_2\in V(G)$.
Let $\scri Q_{v_1,v_2}$ be the con-chain of the pair $\{v_1,v_2\}$ in~$\scri C(G)$.
We can find, in linear time, an embedding preference $\Pi_{v_1v_2}$ of
$\scri Q_{v_1,v_2}$ such that we require exactly $\ins(G,f)$ crossings
to draw a new edge $f=v_1v_2$ into 
\emph{any} plane embedding $G_0$ of $G$ that honors $\Pi_{v_1v_2}$.
\end{theorem}

Any embedding preference $\Pi_{v_1v_2}$ with the properties as described in   
Theorem~\ref{thm:singleedge} will be called an
\emph{optimal embedding preference of} (the con-chain of) the pair $\{v_1,v_2\}$ in $G$.
Note that such an optimal preference $\Pi_{v_1v_2}$ is not necessarily unique
and, furthermore, that optimality of inserting $f$ does not depend on
embedding specifications/preferences at the con-tree nodes 
other than the internal nodes of $\scri Q_{v_1,v_2}$.

\section{MEI Approximation Algorithm}

We now finally have all the technical ingredients to discuss our
new approximation algorithm for the $\MEI(G,F)$ problem
(Definition~\ref{def:MEI}).
The overall idea of the approximation is to suitably combine the individually
computed optimal embedding preferences over all the edges of $F$,
and to prove that not too many conflicts arise, leading to only few
additional crossings as compared to the sum of the individual optimal solutions.
The aforementioned strong locality, achieved in
Definition~\ref{def:embedding-honor}, will be crucial for the algorithm.

\subsection{Coherence, and repairing insertion paths}

Before jumping into a solution of the MEI problem, we have to analyze the
kind of conflicts that arise between optimal embedding preferences
of two distinct pairs $\{v_1,v_2\}$ and $\{w_1,w_2\}$ from~$F$.
Intuitively,
if the con-chains of $\{v_1,v_2\}$ and $\{w_1,w_2\}$ locally traverse 
the con-tree of $G$ in the same way,
then both of them should have ``the same'' optimal individual embedding preferences
there.

Though, the formal finding is not as straightforward as the previous claim
due to two small complications;
first, the situation of R-nodes is more delicate (cf.\
Definition~\ref{def:coherent}), and second,
an optimal embedding preference of a pair is not necessarily unique
and this has to be taken into an account (cf.\ Lemma~\ref{lem:sameprefpaths}).
Recall that our con-paths can be viewed as paths in the decomposition graph,
and this will be our default view in this section.

\begin{defn}[Coherence in con-paths]
\label{def:coherent}
Consider two con-paths $\scri P,\scri P'$ in the decomposition graph of a
connected planar graph~$G$,
and their non-empty intersection $\scri R=\scri P\cap\scri P'$
(which is a subpath by Claim~\ref{clm:con-intersection}).
We say that $\scri P,\scri P'$ are {\em coherent at} $\nu\in V(\scri R)$
if $\nu$ is an inner node of $\scri R$ and the following holds:
if $\mu_1,\mu_2$ are the two neighbors of $\nu$ on $\scri R$
and $\mu_i$, $i\in\{1,2\}$, is an R-node, then also $\mu_i$
is an inner node of $\scri R$.\defend
\end{defn}
We would like to add a small remark on the situation in which $\mu_i$ is, simultaneously, an
end of $\scri R$ and an end of both $\scri P,\scri P'$.
Assume the insertion edges $f,f'$ defining $\scri P,\scri P'$ end in a common
vertex of the skeleton of $\mu_i$. Then, we \emph{could} also define the
con-paths to be ``coherent'' at $\nu$. However, we refrain from doing so as it would
cause unnecessary complications in the proofs.

\begin{lemma}
\label{lem:sameprefpaths}
Let $G$ be a connected planar graph, $\scri C(G)$ the con-tree of~$G$, and $v_1,v_2,w_1,w_2\in V(G)$.
Consider {\em any} optimal embedding preferences $\Pi$ and $\Pi'$ of
$\{v_1,v_2\}$ and $\{w_1,w_2\}$, respectively.
Construct alternative embedding preferences $\Pi''$ of $\{w_1,w_2\}$ as follows:
use the embedding preferences of $\Pi$ for all nodes at which 
$\scri P(\scri Q_{v_1,v_2}),\scri P(\scri Q_{w_1,w_2})$ are coherent, and those 
of $\Pi'$ for all other nodes of $\scri P(\scri Q_{w_1,w_2})$. 
Then, $\Pi''$ is again optimal for $\{w_1,w_2\}$.
\end{lemma}

\begin{proof}
Let $\scri P=\scri P(\scri Q_{v_1,v_2})$ and $\scri P'=\scri P(\scri Q_{w_1,w_2})$
be the considered con-paths.
We iteratively change $\Pi'$ to $\Pi''$. We consider
the nodes $\nu\in V(\scri P)\cap V(\scri P')$ at which
$\scri P,\scri P'$ are coherent one by one,
and claim by induction that at each step (the order of which is
unimportant) the current embedding preference $\Pi''$ is optimal for $\{w_1,w_2\}$.
The claim is trivial if $\nu$ holds the void preference anyhow.

Let $\mu_1,\mu_2$ be the two neighbors of $\nu$ on $\scri R$.
Assume that $\nu$ is a P-node.
By Definition~\ref{def:embedding-pref-wrt}, the embedding preferences of 
$\Pi$ and $\Pi'$ at $\nu$ are the same, since $\mu_1,\mu_2$ are the same neighbors of $\nu$ on both
paths $\scri P,\scri P'$.
Hence, nothing changes at this step.
By the same argument the preference at $\nu$ remains unchanged when $\nu$ is a C- or S-node,
unless one or both of $\mu_1,\mu_2$ is an R-node.

It remains to consider the case that $\nu$ is a C- or S-node and
$\mu_i$ is an R-node for some $i\in\{1,2\}$.
Let $\sigma_i$ be the neighbor of $\mu_i$ other than $\nu$
on $\scri R$ (since $\mu_i$ is not an end by Definition~\ref{def:coherent}).
Let $a_i$ be the element of the skeleton $S_{\mu_i}$ corresponding to $\nu$
and $b_i$ the element of $S_{\mu_i}$ corresponding to $\sigma_i$
($a_i$ is a cut vertex if $\nu$ is a C-node, but
a virtual edge if $\nu$ is an S-node; analogously for $\sigma_i$).
An optimal solution to the single edge insertion
of $\{v_1,v_2\}$  requires to find
a shortest weighted dual path locally in the rigid skeleton $S_{\mu_i}$
from some face $\varphi_i$ incident with $a_i$ to a face $\psi_i$ incident with $b_i$.
This local setting is the same for $\scri P'$ as for $\scri P$,
and so the same dual path from $\varphi_i$ to $\psi_i$ within $S_{\mu_i}$
can be taken also in an optimal solution to the insertion of $\{w_1,w_2\}$.%

So, we get that the embedding preference of $\Pi'$ at a C-node $\nu$ 
(which lists some face of $S_{\mu_i}$ incident with cut vertex $a_i$)
can be changed in $\Pi''$ to that of $\Pi$ at $\nu$
(which lists $\varphi_i$ as the face incident with $a_i$) while preserving optimality.
This is simultaneously done for $i=1,2$ if both $\mu_1,\mu_2$ are R-nodes.
Similarly for an S-node $\nu$;
a choice of one of the two faces of $S_{\mu_i}$ incident with virtual edge
$a_i$ determines the spin of $a_i$, 
which consequently shows that the switching attribute of $\Pi'$ at $\nu$
can be changed to that of $\Pi$ at $\nu$ while preserving optimality.
\qed\end{proof}

\begin{remark}\label{rem:deterministic}
Algorithmically, it is trivial to ensure that
multiple calls to the same dual shortest-path subproblem $\{a,b\}$ within an R-node skeleton 
always give the same solution. This can, e.g., be achieved picking the start vertex 
for the search from $\{a,b\}$ in a deterministic fashion (e.g., based on some arbitrary 
indexing), and using a deterministic algorithm for the dual path search. Using this method
and with respect to Lemma~\ref{lem:sameprefpaths}, one can easily ensure that 
all the optimal embedding preferences computed individually by
Theorem~\ref{thm:singleedge} already agree at any coherent nodes.
\end{remark}

Since we deal with multiple edge insertion in our paper,
we have to handle situations when not all the optimal embedding preferences
of the pairs from $F$ can simultaneously be satisfied in any plane embedding of $G$.
We use the following terminology.
\begin{defn}\label{def:embedding-honor-defect}
Let $\Pi_U$ be embedding preferences of any set $U$ of con-tree nodes.
We say that an embedding $G_1$ {\em honors the preferences $\Pi_U$ with
defect $r$} if there exists a subset $U'\subseteq U$ of size $|U'|=|U|-r$
such that $G_1$ honors (Definition~\ref{def:embedding-honor}) the restriction $\Pi_{U'}$.
\end{defn}

\begin{lemma}
\label{lem:dirtypassesr}
Let $G$ be a connected planar graph, $v_1,v_2\in V(G)$, and $\Pi_{v_1v_2}$ optimal 
embedding preferences of $\{v_1,v_2\}$.
Assume that $G_1$ is an embedding of $G$ such that $G_1$
honors $\Pi_{v_1v_2}$ with defect $r\geq0$.
Then it is possible to draw a new edge $f=v_1v_2$ into $G_1$
with at most \mbox{$\ins(G,f)+r\cdot\lfloor \Delta(G)/2\rfloor$} crossings.
\end{lemma}
Before we prove this lemma, we may informally describe its statement as follows:
For every individual preference on the con-chain of $\{v_1,v_2\}$ that
is not honored by $G_1$ in the sense of Definition~\ref{def:embedding-honor-defect},
we can apply one ``{\em repair operation}'' to $f$ costing at most
$\lfloor \Delta(G)/2\rfloor$ new crossings (over the optimal insertion).

The following is needed in the proof:
\begin{claim}\label{clm:subpreference}
Let $\scri P_{v_1,v_2}$ be the con-path of a pair $\{v_1,v_2\}$.
Consider an internal P-, C-, or S-node
$\sigma\in V(\scri P_{v_1,v_2})$, and let $w$ be a vertex of the skeleton $S_\sigma$.

(a) $\ins(G,v_1w)+\ins(G,wv_2)\leq \ins(G,v_1v_2)$.

(b) Let $\Pi_{v_1v_2}$ be optimal embedding preferences of $\{v_1,v_2\}$
and $\Pi_{v_1w}$ be their restriction to the internal nodes
of $\scri P_{v_1,w}\subsetneq\scri P_{v_1,v_2}$ of the pair $\{v_1,w\}$.
Then, $\Pi_{v_1w}$ is an optimal embedding preference of $\{v_1,w\}$.
\end{claim}
\begin{proof}
Part (a) follows from the fact that there is 
a vertex cut $X\subseteq V(S_\sigma)$ in $G$ with $w\in X$, 
separating $v_1$ from $v_2$:
if $\sigma$ is a C-node (of the cut vertex $w$) then $X=\{w\}$;
otherwise, there is a suitable
2-cut $X=\{w,x\}$ in the skeleton $S_\sigma$---the latter is a cycle or a parallel bunch.
Hence, in any plane embedding $G_0$ of $G$, the
edge $f=v_1v_2$ has to be drawn through a face $\varphi$ incident with $w$,
and the edges $f_1=v_1w,\,f_2=wv_2$ can be drawn along the arc of~$f$
up to this face~$\varphi$, giving an upper bound on
$\ins(G,f_1)+\ins(G,f_2)\leq\ins(G_0,f)$.
Specially, for $G_0$ that minimizes $\ins(G_0,f)$ we get
$\ins(G,f_1)+\ins(G,f_2)\leq\ins(G_0,f)=\ins(G,f)$.

We prove (b) by means of contradiction.
Assume $f$ is drawn optimally (with $\ins(G,f)$ crossings) into a suitable plane embedding
$G_0$ of $G$, and let $f_1$ and $f_2$ be drawn in $G_0$ as in the previous
paragraph with $c_1$ and $c_2$ crossings each.
Then $c_1+c_2=\ins(G,f)$ due to (a) and optimality.
If $\Pi_{v_1v_2}$ restricted to $\scri P_{v_1,w}$ was not optimal, 
then $\ins(G,f_1)<c_1$ would be achieved by some embedding $G_1$ of $G$.
But then, there is a suitable edge order and/or flipping decision 
at $\sigma$ to combine $G_0$ and $G_1$ across the cut $X$ (from the previous paragraph) 
such that inserting $f$ requires no crossings at $S_\sigma$ and consequently only 
$\ins(G,f_1)+c_2<\ins(G,f)$ crossings altogether---a contradiction.
\qed\end{proof}

\begin{proof}[of Lemma~\ref{lem:dirtypassesr}]
Let $\scri P_{v_1,v_2}$ be the con-path (in the decomposition graph of~$G$) 
of the pair $\{v_1,v_2\}$.
We prove the lemma by induction on $r$.
The base case $r=0$ is already established by Theorem~\ref{thm:singleedge}.

We choose the first node $\sigma$ on $\scri P_{v_1,v_2}$
from the $v_1$-end such that $\Pi_{v_1v_2}$ at $\sigma$ is not honored by $G_1$.
Then $\sigma$ is an internal P-, C-, or S-node of $\scri P_{v_1,v_2}$.
Let $w\in V(G)$ be any vertex of the skeleton $S_{\sigma}$,
and $\scri P_{v_1,w},\scri P_{w,v_2}$ be the con-paths of $\{v_1,w\}$
and $\{w,v_2\}$, respectively.
Note that $\scri P_{v_1,w},\scri P_{w,v_2}\subseteq\scri P_{v_1,v_2}$,
and that---by Claim~\ref{clm:subpreference}(b) and symmetry---the corresponding restrictions
$\Pi_{v_1w},\Pi_{wv_2}$ arising from $\Pi_{v_1v_2}$ are optimal embedding preferences
of $\{v_1,w\}$ and $\{w,v_2\}$, respectively.

Consequently, $G_1$ honors $\Pi_{v_1w}$ as whole (i.e. with defect $0$) 
and $\Pi_{wv_2}$ with defect $r-1$.
So $f_1=v_1w$ can be drawn into $G_1$ with $\ins(G,f_1)$ crossings by
Claim~\ref{clm:subpreference}(b) and $f_2=wv_2$ can be drawn into $G_1$ with
at most $\ins(G,f_2)+(r-1)\cdot\lfloor \Delta(G)/2\rfloor$
crossings by the induction assumption.
Altogether, using Claim~\ref{clm:subpreference}(a), $G_1+f_1+f_2$ is drawn with at most
$\ins(G,f)+(r-1)\cdot\lfloor \Delta(G)/2\rfloor$ crossings,
each one occuring on either $f_1$ or $f_2$.
Consider the arc $g$ arising from joining the edge arcs representing $f_1$ and~$f_2$.
We would like to use $g$ to draw $f$ but $g$ passes through the vertex $w$
(where $f_1,f_2$ meet).
By a tiny perturbation of $g$ in a small neighborhood of
$w$ in $G_1$ we can obtain a proper drawing of $G_1+f$ at a cost
of at most $\lfloor \Delta(G)/2\rfloor$ additional crossings 
with (at most half of the) edges incident to~$w$.
This establishes $\ins(G_1,f)\leq\ins(G,f)+r\cdot\lfloor \Delta(G)/2\rfloor$
\qed\end{proof}

\subsection{The approximation algorithm}

\begin{alg}[Solving MEI with additive approximation guarantee]
\label{alg:MEIapx}
Consider an instance of the multiple edge insertion problem $\MEI(G,F)$: 
given is a connected planar graph~$G$ and a set $F$ of $k$ pairs of vertices of
$G$ such that $F\cap E(G)=\emptyset$.
\begin{enumerate}[(1), leftmargin=*]\vspace{-3pt}
\item \label{it:a1}
Build the con-tree $\scri C:=\scri C(G)$.
\item \label{it:a2}
Let $F=\big\{\{u_i,v_i\}:i=1,2,\dots,k\big\}$.
For $i=1,2,\dots,k$, determine the ({\em unique}) con-chain 
of $\{u_i,v_i\}$ in $\scri C$ and the corresponding con-path $\scri P_i$
with the ends $\alpha_i,\beta_i$,
and, independently for each~$i$, call the algorithm of Theorem~\ref{thm:singleedge}
to compute optimal embedding preferences $\Pi_i$ of $\{u_i,v_i\}$
(see Remark~\ref{rem:coherentconsistent} for consistency).

\item \label{it:a2p}
Denote by $p(\nu):=\{i:\nu\in V(\scri P_i)\setminus\{\alpha_i,\beta_i\}\}$
the set of indices of all the pairs from $F$ that have a 
preference (possibly void) at a con-tree node~$\nu$.
For each $\nu$ of $\scri C$, choose (suitably) a subset $p'(\nu)\subseteq p(\nu)$
according to some rules defined later on, see Remark~\ref{rem:ignoredprefs} for details.

\item \label{it:a3}
For $i\in\{1,2,\dots,k\}$,
let $\pi_i(\nu)$ denote the individual preference (if existent) of $\Pi_i$ at $\nu$.
Let $R(\nu):=\{\pi_i(\nu): i\in p'(\nu)\}$
be the multiset of the individual preferences at $\nu$ requested by (only) 
those relevant con-paths that have been selected in step \ref{it:a2p}.
\begin{enumerate}[a)]\vspace{1pt}
\item\label{it:a3:a} For each node $\nu$ of $\scri C$, 
if $R(\nu)=\emptyset$ then set a resulting preference $\pi_\nu$ arbitrarily or void.
Otherwise, choose a preference $\pi_\nu\in R(\nu)$ 
such that $\pi_\nu$ is among the elements with maximum multiplicity in
$R(\nu)$ (a {\em semi-majority choice}).

\item\label{it:a3:b}
Using Lemma~\ref{lem:realizeallprefs} (see below), compute a 
plane embedding $G_0$ of $G$ that honors the embedding preferences 
$\Pi:=\{\pi_\nu: \nu\in V(\scri C)\}$.
\end{enumerate}
\item \label{it:a4}
Independently for each $i=1,2,\dots,k$,
compute (deterministically; see Remark~\ref{rem:deterministicfixedembeddingins}) the \mbox{insertion} path for $\{u_i,v_i\}$ into the
fixed embedding $G_0$.
\end{enumerate}
\end{alg}

We defer a detailed runtime analysis of this algorithm to the end of the
section. First,
we would like to comment on some of the steps in this algorithm:
\begin{remark}[Step \ref{it:a2} of Algorithm \ref{alg:MEIapx}]
\label{rem:coherentconsistent}
Based on Remark~\ref{rem:deterministic}, we can assume that
 our algorithm produces consistent embedding preferences at coherent nodes (for any pair of con-paths)
 without further treatment.
 If, for any reason, one does not want to honor Remark~\ref{rem:deterministic},
 there is a simple algorithmic workaround. For $i=2,3,\dots,k$;
 for any $j<i$ and any node $\nu$ of $\scri C$ such that $\scri P_i,\scri P_j$
 are coherent at $\nu$, change the individual preference of $\Pi_j$ at $\nu$ to that of $\Pi_i$
 at $\nu$. By Lemma \ref{lem:sameprefpaths}, the final so-modified embedding preferences 
 (still denoted by $\Pi_j$) are still optimal for their respective pairs $\{u_j,v_j\}$.
\end{remark}

\begin{remark}[Step \ref{it:a2p} of Algorithm \ref{alg:MEIapx}]
\label{rem:ignoredprefs}
The practical meaning of step \ref{it:a2p} is that we may choose to ``ignore'' some
(or even all) of the individual preferences requested by (some of) the con-paths.
Herein, we will closely discuss two variants. We may choose $p'(\nu)\subseteq p(\nu)$
\begin{itemize}
\item \emph{arbitrarily}, as long as $p'(\nu)\neq\emptyset$ if $p(\nu)\neq\emptyset$; or
\item specifically, to fulfill upcoming Definition~\ref{def:ignoringsimplicpref}.
\end{itemize}
Although in particular the first variant may sound silly, there are actually two good reasons for 
allowing freedom in the choice of $p'(\nu)$.
First, this leaves plenty of room for (heuristic) algorithm
engineering in practical applications, while still providing a firm
approximation guarantee for essentially any somewhat reasonable choice of $p'(\nu)$
(the first variant, Proposition~\ref{prop:icalpbound}). 
Observe that the first variant also covers
the possibility of simply setting $p'(\nu):=p(\nu)$ for all $\nu\in V(\scri C)$,
and also the perhaps most na\"ive choice, picking one arbitrary
$i\in p(\nu)$ and setting $p'(\nu):=\{i\}$.
Second, for a carefully crafted (and still efficient) choice of $p'(\nu)$
we can in fact provide a stronger worst-case guarantee (the second variant, Theorem~\ref{thm:klogk})
than if we considered all the preferences together.
\end{remark}
 
\begin{remark}[Step \ref{it:a3}\ref{it:a3:a} of Algorithm \ref{alg:MEIapx}]
We do not perform any further optimization of the choice of $\pi_\nu$ in the
algorithm here, even though it can be possible that some embedding specification
could be simultaneously good for several distinct individual preferences at~$\nu$
(again, this leaves room for further possibly heuristic algorithm engineering).
The presented semi-majority choice is just right to prove the algorithm's overall approximation ratio.
\end{remark}

\begin{remark}[Step \ref{it:a4} of Algorithm \ref{alg:MEIapx}]
\label{rem:deterministicfixedembeddingins}
By using a deterministic shortest path algorithm in the dual of $G_0$, we
 can trivially ensure that distinct insertion paths do not cross multiple times. If, for some reason, we do not 
 apply a sufficiently deterministic algorithm, we can simply exchange subpaths as a postprocessing
 step, such that in the end all inserted edges cross each other at most once.
\end{remark}

While deferring the lengthy implementation details of step \ref{it:a3} to
Lemma~\ref{lem:realizeallprefs}, we illustrate the underlying idea of 
Algorithm~\ref{alg:MEIapx} with the following simple claim and its corollary
in Proposition~\ref{prop:icalpbound}.
\begin{claim}\label{clm:twoconflicts}
 Consider the setting of Algorithm~\ref{alg:MEIapx}.
 For any $i\not=j\in\{1,\dots,k\}$,
 there are at most two nodes $\nu$ of\/ $\scri C(G)$ such that both $\pi_i(\nu), \pi_j(\nu)$ exist and
 $\pi_i(\nu)\not=\pi_j(\nu)$, i.e., the computed optimal preferences
 $\Pi_i$ and $\Pi_j$ request different individual preferences at~$\nu$.
\end{claim}
\begin{proof}
 Since individual embedding preferences are stored at the corresponding con-path nodes, conflicting
 preferences may only arise on $\scri R:=\scri P_i\cap\scri P_j$, which is a path by 
 Claim~\ref{clm:con-intersection}. We know that 
 $\Pi_i,\Pi_j$ agree at all coherent nodes in $\scri R$ (either due to the
 deterministic algorithm or after applying Lemma~\ref{lem:sameprefpaths}).
 By Definition~\ref{def:coherent} of coherence, possible conflicts could only be at
 (a) the ends of $\scri R$ and (b) nodes neighboring an R-node that is an end of $\scri R$.
 Recall that R-nodes store the void embedding preference, and so at each end of $\scri R$ there can 
 only be one troublesome node, either of type (a) or (b). This establishes the claim.\qed
\end{proof}

\begin{proposition}[Weak estimate]
\label{prop:icalpbound}
Consider a connected planar graph $G$ and a set $F$ of $k$ vertex pairs over $V(G)$.
Let $\inslb(G,F) := \sum_{f\in F} \ins(G,f)$
be the sum of the individual insertion values---an obvious lower bound for\/ $\ins(G,F)$.
If Algorithm~\ref{alg:MEIapx}, in step~\ref{it:a2p}, chooses \emph{arbitrary} 
$p'(\nu)\not=\emptyset$ for each node $\nu$
of $\scri C(G)$ with $p(\nu)\not=\emptyset$, then the result is a plane embedding $G_0$ of $G$ such that
\begin{align}\label{eq:icalpbound}
\ins(G,F)\>\leq\> \ins(G_0,F)\>\leq\> \inslb(G,F)+
	\left(2\left\lfloor\frac{\Delta(G)}2\right\rfloor+1\right)\cdot{k\choose2}
.\end{align}
\end{proposition}
Note that already this short statement establishes the approximation factor
given for the MEI problem in the conference version of this paper
\cite{CHicalp}; herein (Theorem~\ref{thm:klogk}) we will later establish a stronger 
bound as well.

\begin{proof}
We want to show, in the language of
Definition~\ref{def:embedding-honor-defect},
that the sum of defects of $G_0$ honoring each one of the optimal preferences
$\Pi_i$, $i=1,\dots,k$, from the algorithm is at most $2{k\choose2}$. Inequality
\eqref{eq:icalpbound} would then immediately follow from Lemma~\ref{lem:dirtypassesr}
and the fact that the edges of $F$ pairwise cross at most once.

Assume the notation of Algorithm~\ref{alg:MEIapx}.
Let $\hat{G}_0$ be a good enriched embedding for~$\Pi$.
We say a pair $(\mu,i)$, for any $\mu\in V(\scri C)$ and $1\leq i\leq k$, 
forms a \emph{dirty pass} if $\hat{G}_0$ is not good for $\pi_i(\mu)$
(Definition~\ref{def:embedding-honor}).
Obviously, the total number of dirty passes equals the sum of defects of $G_0$.
We show that all the dirty passes of $\hat{G}_0$ can be counted towards
unordered index pairs $\{i,j\}\subseteq\{1,\dots,k\}$ such that
each such pair is ``responsible'' for at most two dirty passes altogether.
Indeed, if $(\nu,i)$ is a dirty pass, then there exists some $j\in\{1,\dots,k\}$ such that
the computed preference at $\nu$ is $\pi_\nu=\pi_j(\nu)\not=\pi_i(\nu)$.
We hence count $(\nu,i)$ towards $\{i,j\}$ and,
by Claim~\ref{clm:twoconflicts}, we already know that this may happen at
most twice for each pair~$\{i,j\}$.
\qed\end{proof}

Finally, we provide a straightforward implementation and a proof of correctness of step
\ref{it:a3}\ref{it:a3:b} in Algorithm~\ref{alg:MEIapx}.

\begin{lemma}
\label{lem:realizeallprefs}
Let $G$ be a connected planar graph and $\scri C=\scri C(G)$ a con-tree of~$G$.
Assume $\Pi=\{\pi_\nu: \nu\in V(\scri C)\}$ is an arbitrary collection of node 
embedding preferences for $\scri C$.
Then there is a plane embedding $G_0$ of $G$ such that $G_0$ honors the preference $\Pi$.
The embedding $G_0$ can be computed in linear time.
\end{lemma}

\begin{proof}
In the first step we fix a plane embedding $H_1$ of each block $H$ of $G$.
If $H$ is trivial, its embedding is already pre-specified (or even unique).
Otherwise $H_1$ is determined by deducing an embedding together with all the spin values.

Let $\scri T_1\subseteq\scri T(H)$ be a subtree of the sSPR-tree of $H$ stored in $\scri C$.
We prove by induction on $|V(\scri T_1)|$ that there exists a good enriched 
embedding $H_1$ of $H$ for $\Pi$ restricted to (the nodes of) $\scri T_1$.
The claim holds for empty $\scri T_1$.
Let $\nu$ be a leaf of $\scri T_1$, and let $H_2$ be a good enriched embedding 
of $H$ for $\Pi$ restricted to $\scri T_1-\nu$.
Consider the type of $\nu$ as in Definition~\ref{def:SPR}:
\begin{enumerate}[a)]
\item If $\nu$ is an S-node, then $H_1:=H_2$ and we only determine the
associated spin values.
Let $\{\mu_1,\mu_2\}$ be the $\pi_\nu$-peers stored at $\nu$, and 
$e_{\mu_1},e_{\mu_2}$ the respective virtual edges in $S_\nu$.
Actually, if $\mu_i$, $i\in\{1,2\}$, is a C-node, we simply refer as
$e_{\mu_i}$ to the corresponding CS-pair.
Up to symmetry, $\mu_2\not\in V(\scri T_1)$. 
If $\mu_1\not\in V(\scri T_1)$ as well, we set the spin of $e_{\mu_1}$ arbitrarily
(while otherwise it has already been set by $\mu_1$).
In any case, we select the spin of $e_{\mu_2}$ to honor the \lswitching/\lnonswitching
label at $\pi_\nu$, and we charge the remaining spins associated with $\nu$
arbitrarily.
\item If $\nu$ is an R-node, it may be that its neighbor $\mu$ in $\scri T_1$
is an S-node such that $\nu$ is one of the $\pi_\mu$-peers of~$\mu$.
If this is the case, then we get $H_1$ by flipping the embedding of $S_\nu$ 
in $H_2$ such that the spin value and switching label specified by $\mu$ are honored. 
Otherwise, let $H_1:=H_2$.
We also set the spin values of the virtual edges in $S_\nu$ accordingly.
\item If $\nu$ is a P-node, then the $\pi_\nu$-peers are S-nodes $\{\mu_1,\mu_2\}$ 
where, up to symmetry, $\mu_2\not\in V(\scri T_1)$.
We arbitrarily set the spin values of all the virtual edges
$e\not=e_{\mu_2}$ of $S_\nu$, except possibly $e=e_{\mu_1}$
if $\mu_1\in V(\scri T_1)$ (then the spin has been set by $\mu_1$).
For $H_1$ we choose an embedding of $S_\nu$, and a spin value of $e_{\mu_2}$, 
such that $e_{\mu_1},e_{\mu_2}$ form a face as required per 
Definition~\ref{def:embedding-honor}, case {\let\bf\relax\ref{def:it:prefP}}.
\end{enumerate}

Now, we have an enriched embedding $G_1$ of $G$ with the property that, 
for each block $H$ of $G$, $G_1$ induces an enriched subembedding $H_1$ 
that is good for $\Pi$ restricted to $\scri T(H)$. 
To obtain the final embedding $G_0$ of $G$, it remains to
modify $G_1$---only at the cut vertices of $G$\,---such that resulting $G_0$ 
is good for $\Pi$ at all the C-nodes of $\scri C$.
Note that, technically working with spherical embeddings,
we can freely choose the outer face of any $H_1\subseteq G_1$ 
in the plane without changing the embedding specification at any node of $\scri T(H)$.

We again proceed by induction on the size of a suitable subtree 
$\scri B_2\subseteq\scri B(G)$ such that there exists
an embedding $G_2$ of $G$ good for $\Pi$ restricted to all the C-nodes
and all the sSPR-trees of $\scri B_2$.
For technical reasons, $\scri B_2$ needs to be {\em special},
meaning that all the leaves of $\scri B_2$ are B-nodes
(which clearly holds true for whole $\scri B(G)$).
The base of the induction is $\scri B_2$ formed by a singleton B-node
of a block~$H$, for which the subembedding $H_0\subseteq G_2:=G_1$
has been fixed above.

In the induction, any arising $\scri B_2$ will contain a C-node $\gamma$, such that if
$\scri B_3\subseteq\scri B_2$ results by removing $\gamma$ and all its adjacent
B-leaves, then $\scri B_3$ is empty or again a special tree.
By the induction assumption, let $G_3$ be an embedding of $G$ that is good
for $\Pi$ restricted to $\scri B_3$.
Let $c\in V(G)$ be the cut vertex represented by~$\gamma$,
and $\{\mu_1,\mu_2\}$ be the $\pi_\gamma$-peers at $\gamma$
where $\mu_i$ belongs to the adjacent sSPR-tree $\scri T(H_i)$, $i=1,2$.

For $i=1,2$, we take a specific face $\varphi_i$ of the skeleton $S_{\mu_i}$: 
for a D-node $\mu_i$, we take its default face; 
for an R-node, $\varphi_i$ is specified by the label in $\pi_\gamma$;
for an S-node, $\varphi_i$ is determined by the spin value of the CS-pair $(c,\mu_i)$.
Let $\varphi_i'$ be the corresponding face of the (sub)embedded block $H_i\subseteq G_3$.
We may assume, up to symmetry, that $\mu_2$ does not belong to an sSPR-tree
held by $\scri B_3$.
We make $\varphi_2'$ the outer face of $H_2$, and
we construct $G_2$ from $G_3$ by rearranging the embedding specification at $\gamma$ (and, thereby, the blocks incident with $c$)
such that $H_2$ occurs inside the face~$\varphi_1'$.
It is clear that $G_2$ is now good also for $\pi_\gamma$.

\smallskip
Finally, following the constructive steps of this proof,
it is routine to verify that the resulting embedding $G_0$ 
can be computed in overall linear time.
\qed\end{proof}

\subsection{Improved approximation guarantee}

We are going to turn the weak approximation guarantee of
Proposition~\ref{prop:icalpbound}
into an asymptotically optimal one---with the additive $O(\Delta\cdot k^2)$ 
term improved down to $O(\Delta\cdot k\log k +k^2)$.
To achieve this goal, we will count the dirty passes of $G_0$,
and so the overall defect,
similarly to the proof of Proposition~\ref{prop:icalpbound}, 
but with respect to a special order of the con-paths such that, at each step, 
we account for roughly at most $O(\log k)$ new ones.
The two crucial ingredients for this approach are the semi-majority choice
of the preferences $\Pi$ in Algorithm~\ref{alg:MEIapx} and the following folklore%
\footnote{This claim is better known in the following formulation:
The intersection graph of subtrees in a tree is chordal and it
contains a so-called simplicial vertex.}
claim.
\begin{claim}\label{clm:simplicial}
Let $T$ be a tree, and $U_i\subseteq T$, $i=1,2,\dots,h$, be an arbitrary
collection of subtrees of~$T$.
Then there exists $j\in\{1,\dots,h\}$ and $u\in V(U_j)$ such that,
for every $i\in\{1,\dots,h\}$, if $U_i$ intersects $U_j$ then~$u\in V(U_i)$.
\end{claim}
\begin{proof}
Consider any graph $H$, subgraph $H'\subseteq H$, and $v\in V(H)$.
Let the {\em distance from $v$ to $H'$} be the minimum distance $d$ between
$v$ and a vertex of $H'$, and the {\em shore of $H'$ from $v$} be the subset
of those vertices of $H'$ having distance $d$ from~$v$.
Clearly, if $H$ is a tree and $H'$ is connected, then the shore of $H'$ must
always be a single vertex $u$ and every path from $v$ to $H'$ contains~$u$.

Choose any vertex $v\in V(T)$ and let $d'$ be the maximum distance from $v$
to any $U_i$, $i\in\{1,\dots,h\}$.
If $d'=0$ then $u:=v$ fulfills the claim for any~$j$.
Suppose $d'>0$ and let some $U_j$, $j\in\{1,\dots,h\}$, 
with the shore $u\in V(U_j)$, be at distance $d'$ from~$v$.
We show that this choice of $u,j$ fulfills the claim.
Consider any $U_i$, $i\in\{1,\dots,h\}$, that intersects $U_j$. 
Clearly its distance from $v$ is at most $d'$.
Let $x\in V(U_i)\cap V(U_j)$ and let $y$ be the shore of $U_i\cup U_j$ from~$v$.
According to the previous paragraph, the path from $v$ to $x$ must contain 
both of $y,u$, and so $u\in V(U_i)$, too.
\qed\end{proof}

We first briefly outline how Claim~\ref{clm:simplicial} can help to improve the
estimate for Algorithm~\ref{alg:MEIapx} over Proposition~\ref{prop:icalpbound}:

\begin{enumerate}[a)]
\item \label{it:logkoutline-a}
For an instance $\MEI(G,F)$,
assume that the decomposition graph $\scri D:=\scri D(G)$ is a tree
(this happens, e.g., when $G$ is $2$-connected).
Let $\scri P_i\subseteq \scri D$, $i=1,\dots,k$, be the con-paths of the
$k$ insertion pairs $\{u_i,v_i\}\in F$.
Then, by Claim~\ref{clm:simplicial}, there is $j\in\{1,\dots,k\}$
such that all of the con-paths hitting $\scri P_j$ do so in the same node
$\nu\in V(\scri P_j)$.
\item 
Let $\scri P_j',\scri P_j''$ be the two half-paths of $\scri P_j$ from
\ref{it:logkoutline-a}
starting in $\nu$, and let $\ell< k$ be the total number of con-paths 
sharing~$\nu$ with $\scri P_j$.
We can claim that the embedding $G_0$ computed in Algorithm~\ref{alg:MEIapx}
honors the preferences $\Pi_j$ restricted to $\scri P_j'$ with defect at most
\mbox{$\log_2\ell<\log_2k$\,}: informally, by our semi-majority choice,
the embedding might not be good only for individual preferences of $\Pi_j$ at those nodes of
$\scri P_j'$ where at least half of all participating con-paths ``divert
from'' $\scri P_j'$ (at this node or at the next R-node).
An upper bound $\log_2k$ for each of $\scri P_j',\scri P_j''$ hence follows
easily, and we may account for $+1$ in the defect sum due to $\nu$ itself.
\item 
If the previous bound was extendable recursively to all the con-paths,
a final estimate for the sum of all the defects would be at most
$1+2\log_2k+1+2\log_2(k-1)+\dots\leq k + 2k\log_2k$
(while the actual bound in Theorem~\ref{thm:klogk} will come out just slightly worse).
\end{enumerate}

For an informal reference, we will call this approach a {\em$\log_2k$-defect argument}.
There are, however, two big problems with the outlined (optimistic) scenario:
\begin{remark}[On utilizing the $\log_2k$-defect argument]
\label{rem:logk-defect-problems}
\begin{enumerate}[a)]
\item\label{it:logk-a}
Since $\scri D(G)$ does not have to be a tree, Claim~\ref{clm:simplicial} 
does not directly hold for it (however, we at least
know that two con-paths may only intersect in a subpath in $\scri D(G)$).
\item\label{it:logk-b}
A somehow less obvious problem with the $\log_2k$-defect argument
pops up when applying it recursively---after removing several of
the con-paths, it is no longer true that a semi-majority choice in this
subcollection is the same as the semi-majority choice made by the algorithm 
(for all the con-paths).
\end{enumerate}
\end{remark}

\begin{figure}[tb]
\begin{minipage}[b]{0.49\textwidth}
\includegraphics[width=\linewidth]{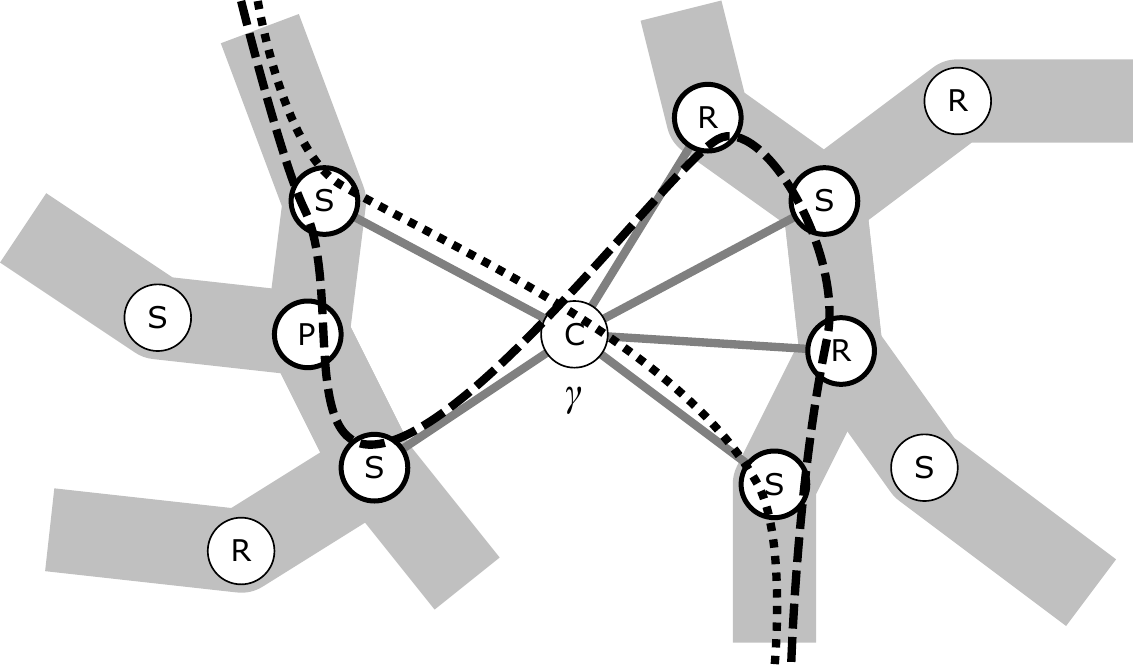}
\end{minipage}\hfill
\begin{minipage}[b]{0.49\textwidth}
(a) We could reroute the dotted con-path to become the dashed con-path, without problems. 
We require no embedding preferences at the newly added nodes, as their skeletons all share the 
cut vertex represented by $\gamma$.
\end{minipage}
\smallskip

\begin{minipage}[b]{0.49\textwidth}
\includegraphics[width=\linewidth]{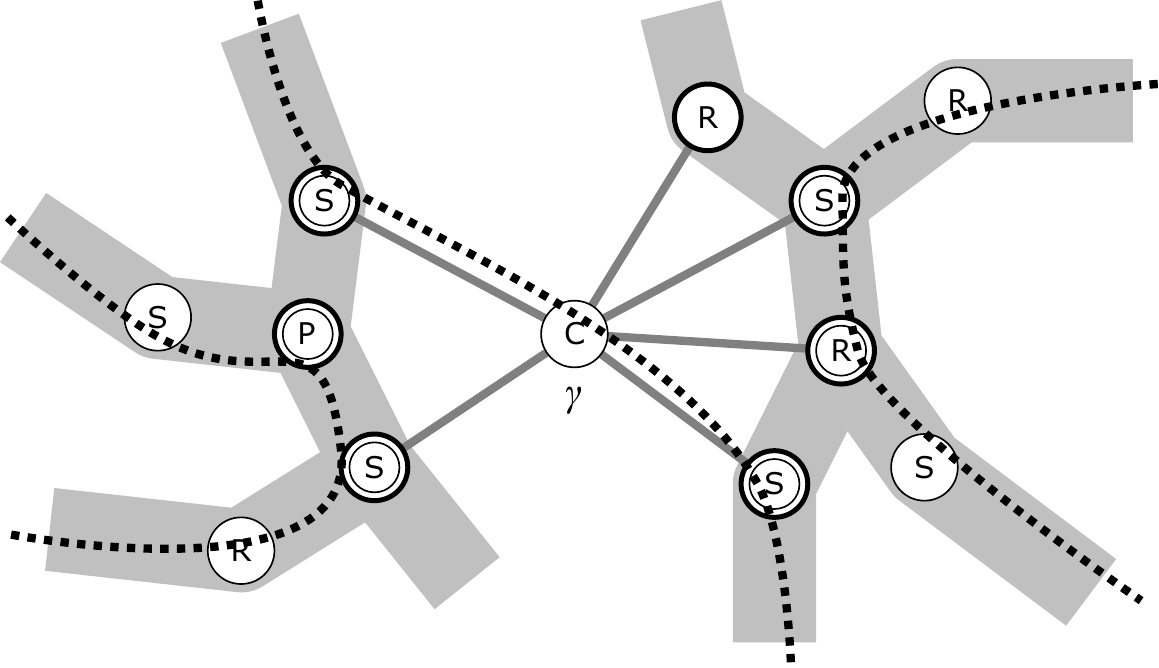}
\end{minipage}\hfill
\begin{minipage}[b]{0.49\textwidth}
(b) For each of the three con-paths, we mark those of their nodes with a double circle 
for which $\gamma$ is a substitute.
\end{minipage}
\smallskip

\begin{minipage}[b]{0.49\textwidth}
\includegraphics[width=\linewidth]{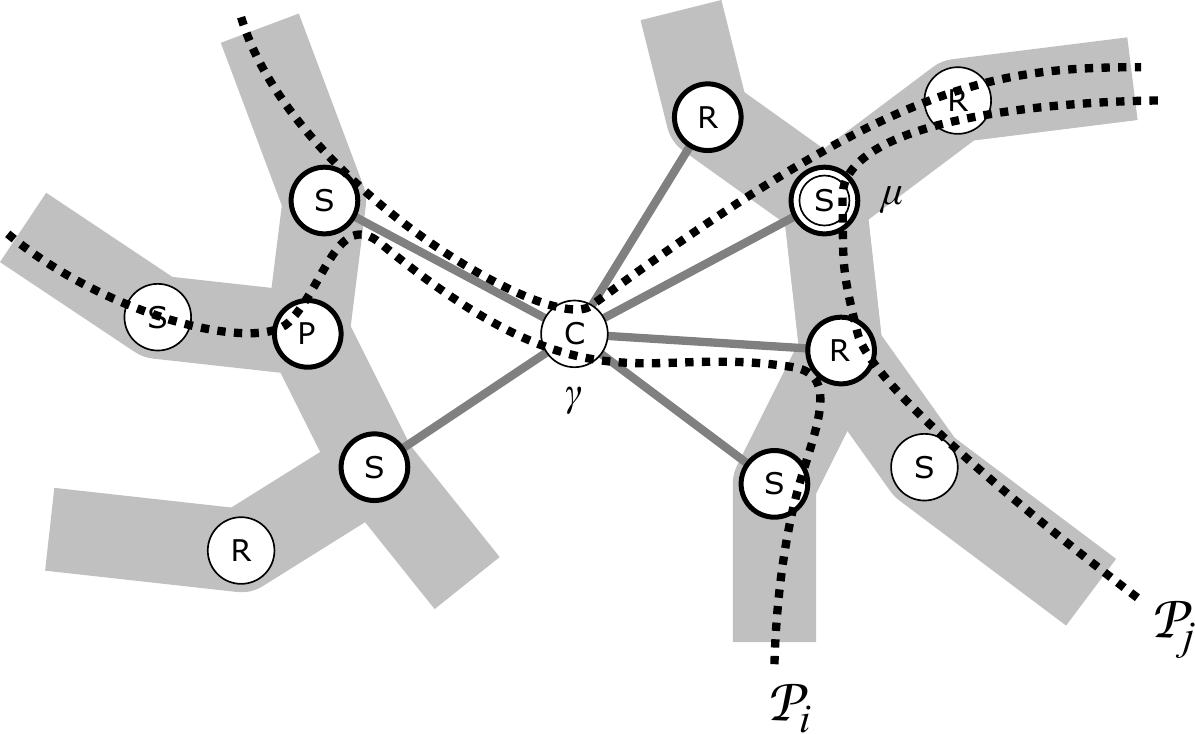}
\end{minipage}\hfill
\begin{minipage}[b]{0.49\textwidth}
(c) In a situation like this, Claim~\ref{clm:simplicial} would fail, as the graph induced 
by the con-paths is not a tree. Lemma~\ref{lem:simplicpath} allows to pick $\scri P_j$ and 
$\mu$ (double circle): for the labeled $\scri P_i$ we have $V(\scri P_i)\cap V(\scri P_j)\neq\emptyset$ 
but it does not contain $\mu$; however, $\gamma\in V(\scri P_i)$ is a substitute for $\mu$ w.r.t.~$\scri P_j$.  
\end{minipage}
\caption{Substitutes w.r.t.\ con-paths. We visualize a part of a decomposition graph $\scri D$, analogously 
to Figure~\ref{fig:exampledecomp}. Con-paths are shown as thick dotted/dashed curves. 
The mates of the central C-node $\gamma$ are marked with thick borders.}
\label{fig:substitute}
\end{figure}

To address Remark~\ref{rem:logk-defect-problems}\,\ref{it:logk-a},
we refine Claim~\ref{clm:simplicial} in following Lemma~\ref{lem:simplicpath}. 
Note, however, that if the given 
graph $G$ would be biconnected, $\scri D(G)$ would be a tree and
we could directly use Claim~\ref{clm:simplicial} and its consequent bound instead.
Consider a con-path $\scri P$ and an internal node $\nu\in V(\scri P)$.
Recall the notion of a mate from Definition~\ref{def:flat-con-tree}, and 
the fact that a mate of a cut vertex $x$ may also be a P-node in which case 
it is not directly adjacent to $x$'s C-node in $\scri D$.
We say that a node $\gamma\in V(\scri D)$ is a
{\em substitute for $\nu$ w.r.t.~$\scri P$} if (cf.\ Figure~\ref{fig:substitute}):
\begin{itemize}
\item $\gamma$ is a C-node of a cut vertex $x$ of~$G$ such that
	$\nu$ is a mate of~$x$ (and so $\nu$ is not a C-node), and
\item there is a neighbor $\mu$ of $\nu$ on $\scri P$ such that $\mu=\gamma$ or 
	$\mu$ is a mate of~$x$, too.
\end{itemize}
In the following lemma, we consider \emph{some} index set $I$. 
While it is most natural to think of $I$ as $\{1,\ldots,k\}$, we will later 
apply this lemma also for subsets of the latter.

\begin{lemma}
\label{lem:simplicpath}
Let $G$ be a connected planar graph, $I$ be a set of indices, and
$\scri P_i$, $i\in I$, be a collection of con-paths in the decomposition
graph $\scri D:=\scri D(G)$ (as derived from an arbitrary collection
of con-chains in $\scri C(G)$).
Then there exists $j\in I$ and $\mu\in V(\scri P_j)$ such that,
for all $i\in I$, the following holds:
if $V(\scri P_i)\cap V(\scri P_j)\not=\emptyset$,
then $\mu\in V(\scri P_i)$ or $V(\scri P_i)$ contains a substitute for $\mu$
w.r.t.~$\scri P_j$.
\end{lemma}
Note that different $\scri P_i$'s in the statement may have different substitutes 
for $\mu$, but there are at most four available substitutes for $\mu$
w.r.t.~$\scri P_j$ anyway (one for each vertex of the two virtual edges of
the neighbors of $\mu$ on $\scri P_j$). 

\begin{proof}
We would like to apply Claim~\ref{clm:simplicial} but $\scri D$ is not a tree
(due to adjacencies of the C-nodes, see Definition~\ref{def:flat-con-tree} and
Figure~\ref{fig:substitute}(c)).
Consider a C-node $\gamma\in V(\scri D)$ of a cut vertex $x$ of $G$.
If $H\subseteq G$ is a block of $G$ incident with~$x$, then the set of all
mates of $x$ in $\scri T(H)$ induces a subtree $\scri T_\gamma(H)\subseteq\scri T(H)$.
We construct a spanning tree $\scri D'\subseteq\scri D$ as follows:
\begin{itemize}
\item For every C-node $\gamma$ of $\scri D$ and each incident block
$H\subseteq G$, we delete all but (an arbitrary) one of the edges of $\scri D$ between
$\gamma$ and $V(\scri T_\gamma(H))$.
\item For each $\scri P_i$, $i\in I$, such that
$\scri P_i$ passes through any such $\gamma$ and one of its deleted edges,
we reroute $\scri P_i$ as $\scri P_i'$ along the remaining edge (between $\gamma$ and
$\scri T_\gamma(H)$) and through $\scri T_\gamma(H)$.
Otherwise, let $\scri P_i':=\scri P_i$.
\end{itemize}

Now we apply Claim~\ref{clm:simplicial} onto $\scri D'$ and the paths $\scri P_i'$,
$i\in I$, obtaining a pair $j$ and $\mu'\in V(\scri P_j')$.
We analyze the situation with respect to $\scri D$ and~$\scri P_j$.
\begin{itemize}
\item
If $\mu'\in V(\scri P_j)$, then we set $\mu:=\mu'$.
Otherwise, $\mu'$ is not a C-node and there is a C-node $\gamma\in V(\scri P_j)$ 
such that $\mu'$ is a mate of the cut vertex of~$\gamma$;
then, we set $\mu$ to be the neighbor of $\gamma$ on $\scri P_j$ corresponding to~$\mu'$ (i.e., $\mu$ is the first node in $\scri P_j$
when traversing $\scri P'_j$ starting at $\mu'$ and away from $\gamma$). 
\item
If, for any $i\in I$, $\mu'\not\in V(\scri P_i')$ then
$V(\scri P_i)\cap V(\scri P_j)=\emptyset$; this is since 
$V(\scri P_i)\subseteq V(\scri P_i')$ by our construction.
(Though, the converse direction is not true in general.)
Consequently, our analysis only has to consider those indices $i\in I$
for which $\mu'\in V(\scri P_i')$ and 
$V(\scri P_i)\cap V(\scri P_j)\not=\emptyset$.
\item
If $\mu\in V(\scri P_i)$ then we are done.
This is true, in particular, when $\mu$ is a C-node.
Hence, in addition to the previous paragraph, we may now assume that $\mu$ is not a
C-node and $\mu\not\in V(\scri P_i)$.
Let $H$ be the block of $G$ such that $\mu\in \scri T(H)$.
\item
If $\mu=\mu'$ then, since $\mu\in V(\scri P_i')\setminus V(\scri P_i)$, 
the path $\scri P_i'$ has been rerouted from original $\scri P_i$ 
at a C-node $\gamma'\in V(\scri P_i)$ such that 
$\mu=\mu'\in V(\scri T_{\gamma'}(H))$.
Let $\tau\in \scri T(H)$ be the neighbor of $\gamma'$ on $\scri P_i$
($\tau$ is the only node of $\scri P_i$ in $\scri T_{\gamma'}(H)$).
Since $\scri P_i$ intersects $\scri P_j$ and $\scri T(H)$ has no cycles,
it is $\tau\in V(\scri P_j)\setminus\{\mu\}$.
Consequently, $\gamma'$ is a substitute for $\mu$ w.r.t.~$\scri P_j$ by definition.
\item
It remains to consider, in addition to the previous, that $\mu\not=\mu'$.
In this case the C-node $\gamma$ has been defined above; we have
$\gamma\mu\in E(\scri P_j)$ and $\gamma$ is a substitute for $\mu$ w.r.t.~$\scri P_j$.
So, if $\gamma\in V(\scri P_i)$ then we are done.
Hence $\mu,\gamma\not\in V(\scri P_i)$, and
(a) $\scri P_i$ intersects $\scri P_j$ in $\scri T(H)$ in a node $\varrho\not=\mu$,
or (b) $\scri P_i$ is disjoint from $\scri P_j$ in $\scri T(H)$ but they
intersect in a C-node $\delta\not=\gamma$ incident with the block~$H$.

Consider case (a);
since the only connection from $\mu'$ to $\varrho$ in $\scri T(H)$ contains~$\mu$ 
(and  $\mu\not\in V(\scri P_i)$), there again has to be a C-node $\gamma'\in
V(\scri P_i)$ such that $\mu',\varrho\in V(\scri T_{\gamma'}(H))$, and consequently 
$\mu\in V(\scri T_{\gamma'}(H))$ and $\gamma'$ is a substitute for $\mu$ w.r.t.~$\scri P_j$, too.
In case (b), 
$\scri P_i'\cap\scri P_j'$ contains a path from $\mu'$ to $\delta$
(since $\delta$ has only one edge to $\scri T(H)$ in $\scri D'$),
while the $\scri P_i$-neighbor of $\delta$ in $\scri T(H)$ belongs to
$V(\scri P_i')\setminus V(\scri P_j)$.
Therefore, $\mu\in V(\scri T_{\delta}(H))$ and so $\mu\delta\in E(\scri P_j)$ by minimality.
Then $\delta$ is a substitute for $\mu$ w.r.t.~$\scri P_j$.
\qed\end{itemize}
\end{proof}

An application of Lemma~\ref{lem:simplicpath} in the aforementioned
$\log_2k$-defect argument brings another elusive problem; namely,
what should be the subpaths $\scri P_j',\scri P_j''$ of selected $\scri P_j$
to which the argument is applied?
The crucial property we need is that every con-path $\scri P_i$ hitting
$\scri P_j'$ does so in the starting node of~$\scri P_j'$.

Informally, for $\scri P_j$ and $\mu$ as in Lemma~\ref{lem:simplicpath}
we let $\scri P_j^o\subseteq\scri P_j$ be the minimal subpath intersecting
all the con-paths not disjoint from $\scri P_j$.
Then $\scri P_j',\scri P_j''$ can be chosen as the subpaths edge-disjoint 
from $\scri P_j^o$ such that $\scri P_j'\cup\scri P_j^o\cup\scri P_j''=\scri P_j$.
However, we also have to consider possible defects (an unbounded number of?)
due to honoring the preferences of $\scri P_j^o$.
Let $\alpha^o,\beta^o$ be the ends of $\scri P_j^o$.
We will show that each of the corresponding skeletons $S_{\alpha^o},S_{\beta^o}$ 
shares a vertex with~$S_\mu$ (formal details given later).
Consequently, only at most two repair operations (at these shared vertices)
are enough for the whole insertion path along $\scri P_j^o$,
while the individual preferences along $\scri P_j^o$ are simply ignored
in the algorithm (Remark~\ref{rem:ignoredprefs}).
This accounts for at most $2\log_2k+2$ dirty passes along whole $\scri P_j$,
as desired.

\smallskip
Finally, addressing Remark~\ref{rem:logk-defect-problems}\,\ref{it:logk-b} is
actually not that difficult, and we informally outline the idea now.
Choose any node $\nu$ of $\scri D(G)$ and focus exclusively on $\nu$
and the con-paths passing through it, in other words,
on the multiset $R(\nu)=\{\pi_i(\nu): i\in p'(\nu)\}$ of (selected) individual 
preferences at $\nu$ from Algorithm~\ref{alg:MEIapx}.
Let $\pi_\nu\in R(\nu)$ be the individual preference chosen in the algorithm.

Without loss of generality we may assume that the repeated application of
Lemma~\ref{lem:simplicpath} selects and removes con-paths in the
order $\scri P_1,\scri P_2,\dots,\scri P_k$.
We define $R_1(\nu)=R(\nu)$ and, for $i\in\{2,3,\dots,k\}$, 
$R_{i}(\nu)=R_{i-1}(\nu)\setminus\{\pi_{i-1}(\nu)\}$ (with multiplicity).
The intuitive problem with a na\"ive voting scheme (that would consider the full sets $p(\nu)$
instead of $p'(\nu)$) is that it would be performed over $R(\nu)$ at each node $\nu$; 
our repeatedly applied counting argument based on Lemma~\ref{lem:simplicpath}, 
however, requires a voting according to the corresponding $R_i(\nu)$ at each step.

This can be seen as follows:
If $\pi_1(\nu)\not=\pi_\nu$ then the multiplicity of $\pi_1(\nu)$ in $R(\nu)$ 
is $\leq\frac12|R(\nu)|$ and, informally, we can afford to pay for
one repair operation at~$\nu$ within the $\log_2k$-defect argument.
Otherwise ($\pi_1(\nu)=\pi_\nu$), no repair operation is needed.

For $i\in\{2,3,\dots,k\}$, when $\pi_i(\nu)=\pi_\nu$ or the multiplicity of $\pi_i(\nu)$ in $R_i(\nu)$ 
is $\leq\frac12|R_i(\nu)|$, the argument is fine again.
Consider the contrary, that $\pi_i(\nu)\not=\pi_\nu$
and so the multiplicity of $\pi_i(\nu)$ in $R(\nu)$ is
$\leq\frac12|R(\nu)|$, while the multiplicity of $\pi_i(\nu)$ in $R_i(\nu)$ is
$>\frac12|R_i(\nu)|$.
This may happen, though, it is an easy exercise to show that
every such $i$ can be accounted towards some $j<i$ such that
$\pi_j(\nu)=\pi_\nu$ and the multiplicity of $\pi_j(\nu)$ in $R_j(\nu)$ is
$\leq\frac12|R_j(\nu)|$.
In the latter case of $j$, the $\log_2k$-defect argument reserved one repair
operation at $\nu$ for $\scri P_j$ but it was not required there.
In the formal proof below, we will say 
that $\scri P_j$ issues a ``repair ticket'' at $\nu$ which is
subsequently used by $\scri P_i$ for an actual repair operation.

\smallskip
We are finally ready to prove our stronger bound.
First, we define what implementation of step \ref{it:a2p} 
of Algorithm~\ref{alg:MEIapx} is needed to get our stronger bound.

\begin{defn}[Ignoring some individual preferences based on a good simplicial sequence]
\label{def:ignoringsimplicpref}
Consider the decomposition graph $\scri D$ of a connected planar graph~$G$,
and con-paths $\scri P_1,\dots,\scri P_k$ in~$\scri D$.
\begin{enumerate}[i)]\item
For any permutation $\sigma$ of the indices,
we say that a sequence $\big((\scri P_{\sigma(1)},\mu_{\sigma(1)}),$ $\dots,
	(\scri P_{\sigma(k)},\mu_{\sigma(k)})\big)$,
where $\mu_j\in V(\scri P_j)$, is a {\em good simplicial sequence}
if the following holds true for $i=1,\dots,k$:
\begin{itemize}\item
for $I=\{\sigma(i),\sigma(i+1),\dots,\sigma(k)\}$, the choice 
$j:=\sigma(i)$ and $\mu:=\mu_{\sigma(i)}$ satisfies the conclusion of
Lemma~\ref{lem:simplicpath}.
\end{itemize}
\item\label{it:whatignore}
Let $p(\nu)$ be defined as in Algorithm~\ref{alg:MEIapx}.
We say that a system of sets $p'(\nu)\subseteq p(\nu)$, $\nu\in V(\scri D)$,
{\em ignores the sequence} $\big((\scri P_{\sigma(1)},\mu_{\sigma(1)}),$
$\dots,(\scri P_{\sigma(k)},\mu_{\sigma(k)})\big)$
if the following holds true for $i=1,\dots,k$ and all $\nu$ of~$\scri D$:
\begin{itemize}
\item
if $i\in p(\nu)$, then $i\not\in p'(\nu)$ if and only if one of the following holds true;
$\nu=\mu_{i}$, or $\nu$ is a C-node neighboring $\mu_{i}$ on $\scri P_i$, or
there is a cut vertex $x$ of~$G$ such
that both $\nu$ and $\mu_{i}$ are mates of~$x$.
\defend\end{itemize}
\end{enumerate}
\end{defn}
Note that a good simplicial sequence always exists by
Lemma~\ref{lem:simplicpath}, but it may not be unique
(which is not a problem for our arguments).
Also observe that our~$p'$, as now defined in Definition~\ref{def:ignoringsimplicpref}, 
``ignores'' many more individual preferences than originally suggested in 
the aforementioned sketch of the proof idea. 
This allows for a simpler definition and does not hurt the argument.
Most importantly, Definition~\ref{def:ignoringsimplicpref}\,\ref{it:whatignore}
is invariant on the order $\sigma$ of the simplicial sequence---%
we only use information that a certain node $\mu_i$ is assigned to $\scri P_i$,
$i\in\{1,\dots,k\}$ when defining $p'$.

\begin{theorem}[Strong estimate, Theorem~\ref{thm:main}(\ref{eq:ourMEI})]
\label{thm:klogk}
Consider a connected planar graph $G$ and a set $F$ of $k$ vertex pairs over $V(G)$.
Let $\inslb(G,F) := \sum_{f\in F} \ins(G,f) \leq\ins(G,F)$.
If step \ref{it:a2p} of Algorithm~\ref{alg:MEIapx} is implemented such that
the computed set system $p'(\nu)$, $\nu\in V(\scri D)$, 
ignores some good simplicial sequence over the con-paths
$\scri P_1,\dots,\scri P_k$, then the algorithm
outputs a plane embedding $G_0$ of $G$ such that
\begin{align}\label{eq:klogkbound}
\ins(G,F)\>\leq\> \ins(G_0,F)\>\leq\> \inslb(G,F)+
	\left\lfloor\frac{\Delta(G)}2\right\rfloor\cdot
		2k\lfloor\log_22k\rfloor	+{k\choose2}
.\end{align}
\end{theorem}

\begin{proof}
We assume all the notation of Algorithm~\ref{alg:MEIapx}.
We may also assume, without loss of generality, that the good simplicial
sequence considered when defining $p'$ is 
 $\big((\scri P_{1},\mu_{1}),$ $\dots,(\scri P_{k},\mu_{k})\big)$.
Recall that the actual order of pairs in this sequence
is significant only for the following analysis and it is not used in any way in
Algorithm~\ref{alg:MEIapx}, when defining $p'$.
Denote by $q_\ell(\nu):=p'(\nu)\cap\{\ell,\ell+1,\dots,k\}$.

For $j=1,\dots,k$ in this order, we argue as follows.
By Definition~\ref{def:ignoringsimplicpref}, $j\not\in p'(\mu_j)$.
Let $\alpha_j,\beta_j\in V(\scri P_j)$ be the ends of $\scri P_j$.
Denote by $\alpha_j^0\in V(\scri P_j)\setminus\{\alpha_j\}$ the node
closest to $\alpha_j$ on $\scri P_j$ and such that $i\not\in
p'(\alpha_j^0)$, and by $\beta_j^0\in V(\scri P_j)\setminus\{\beta_j\}$
the one closest to $\beta_j$ such that $j\not\in p'(\beta_j^0)$.
We observe, by Definition~\ref{def:ignoringsimplicpref}, that the skeleton of $\alpha_j^0$ 
(and analogously $\beta_j^0$) has a common vertex with the skeleton of $\mu_j$. 
Let $\scri P_j',\scri P_j''\subseteq\scri P_j$ denote the subpaths of $\scri P_j$
from $\alpha_j^0$ to $\alpha_j$, and from $\beta_j^0$ to $\beta_j$.
If $\alpha_j=\mu_j$ or $\beta_j=\mu_j$, then $\scri P_j'$ or $\scri P_j''$
is undefined and hence the following argument is simply skipped for it.

Note that $V(\scri P_j')\cap V(\scri P_j'')\subseteq\{\mu_j\}$
if both $\scri P_j',\scri P_j''$ are defined.
The subsequent argument will be given only for $\scri P_j'$, with
understanding that the same symmetrically applies to $\scri P_j''$.

\smallskip
By Lemma~\ref{lem:simplicpath} and the definition of a substitute, 
if a con-path $\scri P_i$, $i>j$, intersects $\scri P_j$, 
then $\scri P_i$ contains $\alpha_j^0$ or $\beta_j^0$,
or $\scri P_i$ intersects $\scri P_j-V(\scri P_j'\cup\scri P_j'')$.
Hence, by Claim~\ref{clm:con-intersection}, if $\scri P_i$ intersects $\scri
P_j'$ then $\alpha_j^0\in V(\scri P_i)$, and so we have
$q_j(\alpha_j^{0})\supseteq q_j(\alpha_j^{1})\supseteq
\dots \supseteq q_j(\alpha_j^a)$ where
$\alpha_j^0,\alpha_j^1,\dots,\alpha_j^a=\alpha_j$ 
denote the nodes of $\scri P_j'$ (of length $a$) in this order.
For $\nu\in V(\scri P_i)$ let 
$\tilde q_\ell^{\,i}(\nu)\subseteq q_\ell(\nu)$ 
denote the subset of those $m\in q_\ell(\nu)$ for which $\scri P_m$ 
is coherent with $\scri P_i$ at~$\nu$ or $i=m$.
We say that node $\nu$ is {\em divergent for $\scri P_i$ on level $\ell$}
if $\nu\in V(\scri P_i)$ is not an R-node and 
$|\tilde q_\ell^{\,i}(\nu)|\leq\frac12|q_\ell(\nu)|$.

Our key observation is that the number of nodes of $\scri P_j'$ that are
divergent for $\scri P_j$ on level $j$, is at most
$\lfloor\log_2|q_j(\alpha_j^0)|\rfloor\leq\lfloor\log_2k\rfloor$: 
whenever a node $\alpha_j^i\in V(\scri P'_j)$ is divergent
we have, by the definition of coherence,
$|q_j(\alpha_j^{i+1})|\leq\frac12|q_j(\alpha_j^{i})|$ if $\alpha_j^{i+1}$ is
not an R-node, and
$|q_j(\alpha_j^{i+2})|\leq\frac12|q_j(\alpha_j^{i})|$ if $\alpha_j^{i+1}$ is 
an R-node (recall that R-nodes are not counted as divergent).
This upper bound of $\leq\lfloor\log_2k\rfloor$ divergent nodes along
$\scri P_j'$ will be used to bound the total defect of $G_0$ in honoring
the computed optimal embedding preferences of edges in~$F$.

Let $\nu\in V(\scri P_j')$.
If $\pi_j(\nu)\not=\pi_\nu$ then, by the definitions and the semi-majority
choice in the algorithm, $\nu$ is divergent for $\scri P_j$ on level~$1$.
Unfortunately, it might happen that $\nu$ is divergent for $\scri P_j$ on
level~$1$ but not divergent for $\scri P_j$ on level~$j$,
and we cannot simply account for the cost of this defect at $\nu$ 
on the same level~$j$.
To resolve such cases, we use an amortized analysis which ``borrows''
for the cost from smaller levels.
This is formalized as follows:
\begin{enumerate}[I)]
\item If $\nu$ is divergent for $\scri P_j$ on level~$j$, then we issue one
{\em repair ticket} to $\nu$ (the total number of tickets issued along
$\scri P_j'$ is thus at most $\lfloor\log_2k\rfloor$, as desired).
\item If $\pi_j(\nu)\not=\pi_\nu$, then we use
one of the repair tickets issued to $\nu$ towards a total defect in honoring 
the optimal preferences of $\scri P_j$ restricted to $\scri P_j'$.
\item
Let $m$ be such that $\pi_m(\nu)=\pi_\nu$ and define
$$r_j := \max\big\{
	|\tilde q_{j+1}^{\,i}(\nu)|-|\tilde q_{j+1}^{\,m}(\nu)|:
		j<i\leq k\big\} \geq0
.$$
After finishing steps (I) and (II) on level~$j$,
the number of available (i.e., issued and not yet used) repair tickets for
$\nu$ is at least~$r_j$.
\end{enumerate}

We claim that this amortized analysis is sound, more precisely,
that every time we would like to use a repair ticket in (II),
there is one available.
In a ``simple'' situation, step (II) would simply use the ticket just issued in
(I) on the same level (this happens, e.g., for~$j=1$).
Assume that no ticket is issued on level~$j$,
which means that $\nu$ is not divergent for $\scri P_j$,
and so $|\tilde q_j^j(\nu)| > \frac12|q_j(\nu)|$.
If $\pi_j(\nu)\not=\pi_\nu$, then 
$|\tilde q_j^{\,m}(\nu)| < \frac12|q_j(\nu)|$ and so
$r_{j-1}\geq|\tilde q_{j}^{\,j}(\nu)|-|\tilde q_{j}^{\,m}(\nu)|>0$.
By (III), we have got an available repair ticket from one of the
previous levels for use on level~$j$.

It is thus enough to prove (III) by induction on $j\geq0$.
For the base case $j=0$ (i.e., before the process starts), we have $r_0=0$
by the semi-majority choice of~$\pi_\nu$.
Further on, let $i>j$ be such that
$r_j=|\tilde q_{j+1}^{\,i}(\nu)|-|\tilde q_{j+1}^{\,m}(\nu)|$.
It suffices to discuss the inductive step in two cases,
either $r_j-r_{j-1}\geq1$ or $r_j=r_{j-1}>0$:
\begin{itemize}
\item
Assume $r_j-r_{j-1}\geq1$.
Since $r_{j-1}\geq|\tilde q_{j}^{\,i}(\nu)|-|\tilde q_{j}^{\,m}(\nu)|$
and $\tilde q_{j+1}^{\,i}(\nu)\subseteq\tilde q_{j}^{\,i}(\nu)$,
we have $|\tilde q_{j+1}^{\,m}(\nu)|=|\tilde q_{j}^{\,m}(\nu)|-1$.
Then $\pi_j(\nu)=\pi_m(\nu)=\pi_\nu$ and $\nu$ is divergent for $\scri P_j$
on level~$j$.
Consequently, the repair ticket issued in (I) on level $j$ is not used in
(II) and, indeed, there are at least $r_{j-1}+1=r_j$ available repair
tickets afterwards.
\item
Assume that $r_j=r_{j-1}>0$ and a repair ticket is used in (II) on
level~$j$ since $\pi_j(\nu)\not=\pi_\nu$.
Then $\tilde q_{j+1}^{\,m}(\nu)=\tilde q_{j}^{\,m}(\nu)$ and so
$|\tilde q_{j+1}^{\,i}(\nu)|=|\tilde q_{j}^{\,i}(\nu)|$.
Consequently, $j\not\in\tilde q_{j}^{\,i}(\nu)$ and hence $\nu$ is
again divergent for $\scri P_j$ on level~$j$.
This means a repair ticket for (II) has just been issued in (I) on level~$j$.
\end{itemize}
Claim (III) is finished.

Altogether, we have issued at most $\lfloor\log_2k\rfloor$ repair tickets
along $\scri P_j'$, and also at most $\lfloor\log_2k\rfloor$ tickets along
$\scri P_j''$.
Furthermore, two special repair tickets are issued to $\mu_j$, summing up to
at most $2\lfloor\log_2k\rfloor+2=2\lfloor\log_22k\rfloor$ for iteration~$j$.
For the whole problem, altogether at most $2k\lfloor\log_22k\rfloor$ 
repair tickets are issued.
We are hence nearly finished and it remains to prove that the above distributed
repair tickets are sufficient for all the ``repair operations'' carried out
when drawing the edges of $F$ into~$G_0$.

\smallskip
Again for $j=1,\dots,k$ (the order is now irrelevant), we argue as follows.
By Definition~\ref{def:ignoringsimplicpref}, the skeletons of $\alpha_j^0$ and
$\mu_j$ must share a vertex $x_j\in V(G)$ ($x_j$ is a cut vertex of $G$ if
$\mu_j\not=\alpha_j^0$ but this is not important now).
We analogously set $y_j\in V(G)$ on the side of $\beta_j^0$.
Now, $\scri P_j'$ is the con-path of $\{u_j,x_j\}$ and $\scri P_j''$ is the 
con-path of $\{y_j,v_j\}$.
By Claim~\ref{clm:subpreference}(b), the preferences $\Pi_j'$, which are the restriction of
$\Pi_j$ to $\scri P_j'$, are optimal embedding preferences of $\{u_j,x_j\}$.
The embedding $G_0$ hence honors $\Pi_j'$ with defect at most equal to the number of
repair tickets used along $\scri P_j'$\,---we denote this number by $t_j'$.
We define $\Pi_j''$ and $t_j''$ analogously.

By Lemma~\ref{lem:dirtypassesr}, we can draw the new edges $u_jx_j,y_jv_j$
into $G_0$, constructing $G_1:=G_0+u_jx_j+y_jv_j$ with 
\mbox{$\inslb(G,\{u_jx_j,y_jv_j\})+(t_j'+t_j'')\cdot\lfloor \Delta(G)/2\rfloor$}
crossings.
Furthermore, the new edge $x_jy_j$ can be drawn into $G_1$ with
$\ins(G_1,x_jy_j)$ crossings---this may actually be a nonzero number if
$\mu_j$ is an R-node, but there are no embedding preferences for such 
R-nodes anyway, as their skeleton embeddings are not mutable.
Altogether, $G_2:=G_1+x_jy_j$ is drawn with 
\mbox{$\inslb(G,\{u_jx_j,x_jy_j,y_jv_j\})+(t_j'+t_j'')\cdot\lfloor
	 \Delta(G)/2\rfloor$} crossings.

Using the same argument as in the proof of Lemma~\ref{lem:dirtypassesr},
we now perturb the path formed by the edges $u_jx_j,x_jy_j,y_jv_j$ in $G_2$
into a drawing of the edge $f_j=u_jv_j$ in $G_0+f_j$.
For this, we require at most $2\cdot\lfloor \Delta(G)/2\rfloor$ additional
crossings. The number of crossings in $G_0+f_j$ is hence at most
$\ins(G,f_j)+(t_j'+2+t_j'')\cdot\lfloor\Delta(G)/2\rfloor$
using Claim~\ref{clm:subpreference}(a).

While neglecting the at most $k\choose2$ crossings between the edges of $F$,
we get that the total number of crossings between $G_0$ and $F$ is at most
\begin{align*}
\inslb(G,F)+\sum_{j=1}^k
	(t_j'+2+t_j'')\cdot\left\lfloor\frac{\Delta(G)}2\right\rfloor
\leq
\inslb(G,F)+\left\lfloor\frac{\Delta(G)}2\right\rfloor\cdot
		2k\lfloor\log_22k\rfloor
.\end{align*}
This concludes the proof.
\qed\end{proof}

\subsection{Tightness of the Analysis of Algorithm~\ref{alg:MEIapx}}

Reading the fine-grained analysis of Algorithm~\ref{alg:MEIapx} in
Theorem~\ref{thm:klogk}, it is natural to think whether perhaps the ideas
can be improved further, giving an even tighter guaranteed relation
between the outcome of the algorithm and the sum of the individual insertion
values $\inslb(G,F)$ than \eqref{eq:klogkbound}.
This is, however, not possible as we now show by exhibiting an
asymptotically matching lower bound in Proposition~\ref{prop:tight}.

\begin{proposition}\label{prop:tight}
 For any integers $r,k\geq0$ and~$\Delta\geq4$,
 there exist instances of the $\MEI(G,F)$ problem such that
 $k=|F|$, $\Delta(G)\leq\Delta$, $\inslb(G,F)\geq r$, and
 \begin{equation}\label{eq:lowerapprox}
 \ins(G,F)\>\geq\> \inslb(G,F) + \Omega\!\left(
		\Delta\cdot k\log_2 k + {k\choose 2} \!\right)
 \end{equation}
where $\inslb(G,F) := \sum_{f\in F} \ins(G,f)$.
\end{proposition}

Note, though, that this lower bound only concerns the relation between the
optimum value $\ins(G,F)$ and the simple lower bound $\inslb(G,F)$ we use in our
analysis; it does not say anything about approximability (or
inapproximability) of the MEI problem itself.
The main message of this claim hence is that if one wants to achieve an
algorithm with a tighter approximation guarantee for the MEI problem, then
one must consider something more than just the individual insertion
solutions.

\begin{figure}[tb]
\subfigure[In this picture $\ell\choose2$ crossings are {required}.\label{fig:tight-choose}]{\includegraphics[scale=0.7]{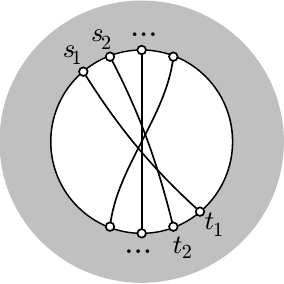}}\hfill
\subfigure[Graphs $H^4$ and $H^6$; there are $4$ or $6$ marked vertices, $2m$ disjoint (concentric) inner cycles,
	and each depicted line represents $\Delta/4$ parallel edges.\label{fig:tight-Hs}]{\includegraphics[scale=1]{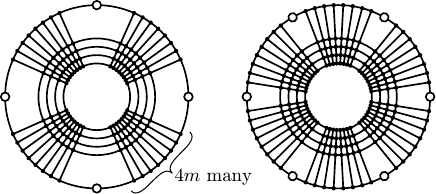}}
\\[1ex]
\subfigure[A detail of R-node gadgets for (d), made of a copy of $H^6$ and $H^4$.\label{fig:tight-Rgadget}]{\includegraphics[scale=1]{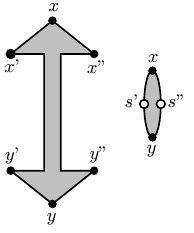}}\hfill
\subfigure[A scheme of the construction requiring $\frac12\Delta\cdot m\log_2m$
	crossings. Each shaded part in the picture is a copy of
	$H^4$ or $H^6$ with the depicted marked vertices.\label{fig:tight-R}]{\includegraphics[scale=1]{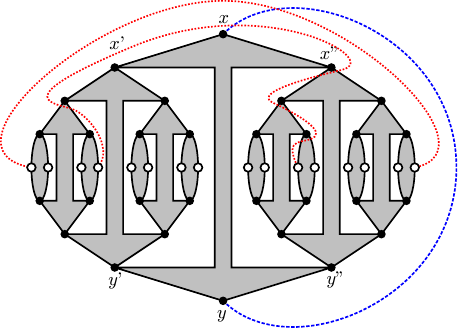}}
 %\caption{Tightness of the approximation algorithm's upper bound (see proof of Claim~\ref{clm:tight}).}
 \caption{Illustration of the proof of Proposition~\ref{prop:tight}:
	constructions requiring many crossings in an optimum MEI solution
	compared to individual edge insertions.}
 \label{fig:tight}
\end{figure}

\begin{proof}
We are going to present three separate constructions. They can then be easily
combined together by adjusting them appropriately to~$k$
and adding dummy edges between them (to satisfy connectivity of~$G$):
\begin{enumerate}[I)]
 \item There exists a planar graph $G^1$ and a vertex pair $a,b\in V(G^1)$
  such that $\ins(G^1,ab)\geq r$ and $\Delta(G^1+ab)=3$.
 \item There exists a planar graph $G^2$ and a set of $\ell$ vertex pairs
  $F_2=\{a_ib_i: a_i,b_i\in V(G^2), i=1,2,\dots,\ell\}$, such that each $G^2+a_ib_i$ is
  planar, $\Delta(G^2)\leq4$, and $\ins(G^2,F_2)={\ell\choose2}$.
 \item There exists a planar graph $G^3$ and a set of $m$ vertex pairs
  $F_3=\{s_it_i: s_i,t_i\in V(G^3), i=1,2,\dots,m\}$, such that each $G^3+s_it_i$ is
  planar, $\Delta(G^3)\leq\Delta$,
  and $\ins(G^3,F_3)\geq\frac12\Delta\cdot m\log_2m$.
\end{enumerate}

 In (I), we take as $G^1$ a sufficiently large plane hexagonal grid, and
 select two edges in $G^1$ that are sufficiently far from each other and from the
 grid boundary; the vertices $a,b$ then subdivide the selected edges.

 In (II), we assume $G^2$ to be any dense and large enough triconnected
 graph with a face of length $2\ell$.  Labeling the vertices on this face
 $s_1,\ldots s_\ell,t_1,\ldots t_\ell$ in clockwise order, gives an instance that
 clearly requires ${\ell\choose 2}$ crossings.
 See Figure~\ref{fig:tight-choose} for an illustration.

 We concentrate on the most interesting construction (III).
 Since the bound in \eqref{eq:lowerapprox} is of asymptotic nature, we may
 without loss of generality assume that $\Delta$ is even and divisible
 by~$4$, and that $m=2^d$ for some integer $d\geq1$.

 Let $H^q$, $q\in\{4,6\}$, be any dense enough planar graph with
 $q$ special \emph{marked} vertices $v_1,\dots,v_q$, each of degree exactly $\Delta/2$, such
 that the following conditions hold.
 Up to mirroring and reordering of parallel edges, $H^q$ allows only a unique plane embedding of $H^q$. 
 Every such embedding $H^q_0$ has a unique face, called
 \emph{active}, incident with all of $v_1,\dots,v_q$ in this order;
 no other face of $H^q_0$ is incident with more than one of $v_1,\dots,v_q$.
 Moreover, assume we embed new \emph{bar} edges $v_1v_4,v_2v_3,v_5v_6$
 (only $v_1v_3$ if $q=4$) into the active face of $H^q_0$; this divides the active
 face into $4$ (or $2$) \emph{sectors}.
 We require that if we draw any curve $\gamma$ with the ends in distinct sectors
 and not crossing the bar edges, then $\gamma$ makes at
 least $\Delta/2$ crossings with the edges of $H^q_0$.
 If the two sectors holding the ends of $\gamma$ are not next
 to each other and $\gamma$ avoids all other sectors,
 then $\gamma$ makes at least $m\Delta$ crossings.

 The desired properties of $H^q$ are for instance achieved by the graphs depicted in
 Figure~\ref{fig:tight-Hs}, where the marked vertices are drawn white, the
 outer face is the active face, and each drawn line represents a bunch of
 $\Delta/4$ parallel edges.

 Our recursive construction of $G^3$ for (III) is schematically depicted in
 Figure~\ref{fig:tight-R}.
 Recalling $m=2^d$, we define a graph $G^s_d$ with two special vertices called the
 {\em poles}, by induction on $d$.
 In the base case, $d=1$, $G^s_1$ is a copy of $H^4$ with the marked
 vertices $x,s'',y,s'$ in this order around the active face, such
 that $x,y$ are the poles of $G^s_1$.
 Having constructed $G^s_d$ and its disjoint copy $G^t_d$, we define
 $G^s_{d+1}$ as follows:
 take a copy $B$ of $H^6$ (the {\em bolt\/}) with the marked vertices $x,x'',y'',y,y',x'$ in this 
 order (see Figure~\ref{fig:tight-Rgadget}), make $x,y$ the new poles of $G^s_{d+1}$,
 and identify $x',y'$ with the two poles of $G^s_d$ and $x'',y''$ with the two poles of $G^t_d$.

 The graph $G^s_d$ has $2^d=m$ marked vertices that are copies of $s',s''$
 from~$G^s_1$ (the white terminals in Figure~\ref{fig:tight-Rgadget}),
 and we denote them by $s_1,\dots,s_m$ (in any order).
 We analogously denote by $t_1,\dots,t_m$ the corresponding vertices in $G^t_d$.
 Finally, we let $G^3:=G^s_{d+1}$ and $F_3=\{s_it_i: i=1,\dots,m\}$.

 It should be understood that the embedding possibilities for $G^3$ are
 essentially defined by the binary flipping decisions of each R-node skeleton
 corresponding to a copy of $H^4$ or $H^6$ in the construction.
 Furthermore, for each $1\leq i\leq m$, the edge $s_it_i$ can be inserted
 into (some embedding of) $G^3$ without crossings
 and into any fixed embedding of $G^3$ with at most $d\Delta$ crossings.
 See the red dotted edges depicted in Figure~\ref{fig:tight-R}.
 To settle (III) it remains to argue that
 $\ins(G^3,F_3)\geq\frac12\Delta\cdot md$, which we achieve in two steps.
 Let $\ins'(G^3,F_3)$ denote the solution value of the $\MEI(G^3,F_3)$
 problem without counting the crossings between edges of $F_3$.

 \begin{claim}\label{clm:deltahalf}
 Let $G^{3+}$ be the graph obtained from $G^3$ by adding a bunch $F_+$ of
 $\Delta/2$ parallel edges between the two poles of $G^3$
 (cf.~the blue dashed curve in Figure~\ref{fig:tight-R}).
 Then $\ins'(G^{3+},F_3)=\ins'(G^3,F_3) +m\Delta/2$.
 \end{claim}
 We have $\ins(G^{3+},F_3)\leq \ins(G^3,F_3) +m\Delta/2$ since both poles of $G^3$ share
 a common face $\varphi$ in the corresponding optimal solution and each edge of $F_3$ traverses 
 any face of embedded $G^3$ at most once. Routing the edges of $F_+$ via $\varphi$ hence gives 
 at most $\Delta/2$ additional crossings per edge of $F_3$.
 Conversely, consider an optimal solution of value $\ins'(G^{3+},F_3)$ and an edge
 $s_it_i\in F_3$.
 We claim that after removing $F_+$ from this solution, we ``save'' at least
 $\Delta/2$ crossings on $s_it_i$.

 If $s_it_i$ crossed all of $F_+$, then we are done.
 Otherwise, by the property of $H^6$, the drawing of $s_it_i$ has to enter
 both the sectors of the bolt of $G^3$ incident to the poles (and so has
 additional $\Delta/2$ crossings in between), or there are
 at least $m\Delta>d\Delta+\Delta/2$ crossings on $s_it_i$.
 In both cases one requires by at least $\Delta/2$ less crossings on $s_it_i$ in
 the subembedding of $G^3$.
 
 \begin{claim}
 $\ins'(G^3,F_3)\geq\frac12\Delta\cdot md$ and
 $\ins'(G^{3+},F_3)\geq \frac12\Delta\cdot m(d+1)$.
 \end{claim}
 By Claim~\ref{clm:deltahalf}, it is enough to prove one of the
 inequalities but it is convenient to consider both of them by indunction
 on~$d\geq0$.
 The base case of $G^3=G^s_1$ ($d=0$) is a degenerate extension of the previous,
 defining $s_1,t_1$ to be the two opposite marked vertices $s',s''$ of
 $H^4$, and it is trivial.

 Stepping from $d-1$ to $d\geq1$; we consider $G^3=G^s_{d+1}$ and an optimal solution
 of value $\ins'(G^{3},F_3)$ inserting the $m$ edges of $F_3$ into a plane embedding
 $G_1$ of $G^3$, without counting crossings inside~$F_3$.
 Let $\beta$ be a curve drawn between the poles of $G_1$ without crossing
 its edges (e.g., the blue dashed curve in Figure~\ref{fig:tight-R}).
 Similarly as in the proof of Claim~\ref{clm:deltahalf}, we can show that
 there is an optimal solution in which every edge of $F_3$ has to cross~$\beta$.

 Let $G_0'$ and $G_0''$ be the subembeddings of the two recursive copies
 of $G^s_d$ in~$G_1$.
 We cut every curve of the edges of $F_3$ at a point in the intersection with $\beta$, 
 to obtain $2m$ curve parts, $m$ of which are incident to $G_0'$. 
 These latter parts can be paired such that the pairs (after reconnection
 along $\beta$) form a set of $m/2$ edges $F_3'$ with ends in $V(G_0')$ and
 the instance $(G_0',F_3')$ is as in the inductive assumption for~$d-1$.
 By the property of $H^6$, the bolt of $G_1$ can play the role of $F_+$ in
 Claim~\ref{clm:deltahalf}; let ${G'_0}^{\!+}$ denote $G_0'$ augmented with the bolt.
 We hence have that the curve parts of $F_3$ incident to $G_0'$ have at least
 $\ins'({G'_0}^{\!+},F_3')$ crossings with edges of $G_1$ by our (second claim
 of the) inductive assumption.
 Since the same symmetrically holds for the curve parts incident to $G_0''$,
 summing together we get that
  $\ins'(G^{3},F_3)\geq2\cdot\ins'({G'_0}^{\!+},F_3')\geq
	2\cdot\frac12\Delta\cdot m/2\cdot(d-1+1)=\frac12\Delta\cdot md$, as desired.
\qed\end{proof}

\begin{remark}
Notice that the actual MEI instances constructed in the proof of
Proposition~\ref{prop:tight} are not at all ``bad'' for us---they
will always be solved to optimality in Algorithm~\ref{alg:MEIapx}.
The example is thus if purely theoretical nature.
\end{remark}

\subsection{Runtime of Algorithm~\ref{alg:MEIapx}}\label{sec:runtime}

Thanks to using well-known building blocks, the overall runtime bound of our algorithm 
is actually rather simple to see.
Let $V=V(G)$.
As mentioned in Section~\ref{sec:prelim}, we can build the con-tree (step~\ref{it:a1})
in linear time $O(|V|)$, based on the linear-time decomposition algorithm~\cite{HT73}.
Recall that $|V(\scri C)|=O(|V|)$.
In step~\ref{it:a2}, we call the (deterministic) $O(|V|)$ insertion algorithm $k$ times.

In step~\ref{it:a2p}, computing $p(\nu)$ for all $\nu\in V(\scri C)$ is trivial in $O(k|V|)$, as is 
achieving the weak estimate by setting, e.g., $p'(\nu)=p(\nu)$. To achieve the strong estimate, we
can compute a simplicial sequence (Claim~\ref{clm:simplicial}) 
via $k$ BFS-traversals in $O(k|V|)$ time. Thereby, we 
also identify node substitutes (Lemma~\ref{lem:simplicpath}) and
decide which preferences to ignore (Definition~\ref{def:ignoringsimplicpref}).

In step~\ref{it:a3}, we can compute $\pi_\nu$ for each of the con-tree nodes $\nu$ via 
semi-majority vote in $O(k)$ time. Computing the embedding $G_0$ based on these
preferences takes only linear time $O(|V|)$. Hence, also this step requires $O(k|V|)$ time.

In step \ref{it:a4}, we then run $k$ (deterministic) BFS algorithms, requiring $O(|V|)$ time each.
Since each edge has at most $O(|V|+k)$ crossings in the end, the realization may require
up to $O(k|V(G)|+k^2)$ time, which hence constitutes the 
overall runtime bound of the algorithm, as given in Theorem~\ref{thm:main}.

\section{Crossing Number Approximations}\label{sec:crApprox}

Our main concept of interest is the crossing number of the graph $G+F$.
We can combine our above result with a result of~\cite{apex}, connecting the optimal
crossing number with the problem of multiple edge insertion.

\begin{thm}[Chimani et al.\ \cite{apex}]
\label{thm:apexrel}
Consider a planar graph $G$ and an edge set $F$, $F\cap E(G)=\emptyset$.
The value $\ins(G,F)$ of an optimal solution to $\MEI(G,F)$ satisfies
\begin{equation*}%\label{eq:insapproxcr}
\ins(G,F) \>\leq\> 2|F|\cdot
	\left\lfloor\frac{\Delta(G)}2\right\rfloor\cdot \crg(G+F)+\binom{|F|}{2}
\end{equation*}
where $\crg(G+F)$ denotes the (optimal) crossing number 
of the graph $G$ including the edges~$F$,
and $\binom{|F|}{2}$ thereby accounts for crossings
between the edges of~$F$.
\end{thm}

Notice that, when considering the crossing number problem of $G+F$, we may
assume $G$ to be connected---otherwise we could ``shift'' some edges of $F$
to~$G$.
Let $k=|F|$, $\Delta=\Delta(G)$.
Plugging the estimate of Theorem~\ref{thm:apexrel} into the place of $\inslb(G,F)\leq\ins(G,F)$
in Theorem~\ref{thm:klogk}, and realizing that the $k\choose2$
term in both estimates stands for the same set of crossings,
we immediately obtain
\begin{align*}
\insapx(G,F) \leq 2k\cdot
        \left\lfloor\Delta/2\right\rfloor\cdot
		 \crg(G+F)+
	2k\lfloor\log_2k\rfloor\lfloor\Delta/2\rfloor
	+{k\choose2}
\end{align*}
Hence we can give the outcome of Algorithm~\ref{alg:MEIapx} as an
approximate solution to the crossing number problem on $G+F$,
proving:

\begin{thm}[Theorem~\ref{thm:main}(\ref{eq:ourCR})]\label{thm:mainCR}
Given a planar graph $G$ with maximum degree $\Delta$ and an edge set $F$,
$|F|=k$, $F\cap E(G)=\emptyset$,
Algorithm~\ref{alg:MEIapx} computes, in $O(k\cdot|V(G)|+k^2)$ time,
a solution to the $\crg(G+F)$ problem with the following number of crossings
$$
\crgapx(G+F) \leq \lfloor\Delta/2\rfloor\!\cdot 2k\!\cdot\crg(G+F)
	+ 2k\,\lfloor\log_2k\rfloor\cdot\left\lfloor\Delta/2\right\rfloor
	+\mbox{$\frac12$}\big(k^2-k\big) .
\vspace*{-1.5ex}$$
\qed\end{thm}

\paragraph{A note on approximating the crossing number of surface-embedded graphs.}
In~\cite{surfaceApprox},
an algorithm is presented to approximate the crossing number of graphs embeddable 
in any fixed higher orientable surface. This algorithm
lists the technical requirement that $G$ has a ``sufficiently dense''
embedding on the surface.
Yet, as noted in~\cite{surfaceApprox}, a result like
Theorem~\ref{thm:mainCR} allows to drop this requirement:
If the embedding density is small, then the removal of the offending small
set(s) of edges is sufficient to reduce the graph genus,
while the removed edges can be later inserted into an intermediate planar
subgraph of the algorithm.% \cite{surfaceApprox}.

\section{A Note on the Planarization Heuristic and the Practicality of our Algorithm}\label{sec:heuristic}

The currently practically strongest heuristic~\cite{GM04} for the crossing number 
problem is the \emph{planarization heuristic} which starts with a maximal
planar subgraph of the given non-planar graph, and then iteratively performs
single edge insertions. The crossings of such an insertion are then replaced by 
dummy nodes such that each edge is inserted into a planar graph. Due to its
practical superior performance, often giving the optimal solution~\cite{mchDiss,carstenDiss}, 
it was an open question if this approach unknowingly guarantees some approximation 
ratio.

By investigating our strategy and proofs, it becomes clear that this approach 
as such cannot directly give an approximation guarantee: by routing an edge 
(in an R-node) through another virtual edge (representing a subgraph $S$) and 
replacing the crossings with dummy nodes, you essentially \emph{fix} (most of) 
the embedding of $S$. This fix might result in $O(n)$ embedding restrictions
for further edge insertions, without having an edge in $F$ that requires this 
embedding. Therefore the
number of dirty passes can no longer be bounded by a function in~$k$.

Yet, an implementation realizing the planarization heuristic already contains 
all the ingredients to obtain our approximation; one ``only'' has to compute 
all the embedding preferences and merge them according to 
Algorithm~\ref{alg:MEIapx}, steps \ref{it:a2p}~and~\ref{it:a3}, before running the
fixed-embedding edge insertion subalgorithm for all inserted edges.
In fact, such an implementation is described in~\cite{DBLP:journals/jgaa/ChimaniG12}, 
where, in a nutshell, it is shown that this algorithm is much 
faster then the best postprocessing-heavy planarization-based heuristics, while 
producing roughly equally good solutions.

\section{Conclusions}

We have presented a new approximation algorithm for the multiple edge insertion problem
which is faster and simpler that the only formerly known one
\cite{sodamultiedge}, while at the same
time giving better bounds; in fact, in contrast to the former multiplicative
approximation, it is the first one with an additive bound. Our algorithm
directly leads also to improved approximations (even with constant ratio 
for a large class of inputs) for the crossing number
problem of graphs in which a given set of edges can be removed in order to obtain a
planar subgraph, and for graphs that can be embedded on a surface of some
fixed genus.

We conclude with an interesting open problem. We know that multiple edge insertion
is NP-hard when the number of inserted edges is part of the input, and it is linear time solvable
for the special case of inserting a single edge. What is the complexity of
optimally inserting a \emph{constant} number of edges?

\bibliography{apxmei}
\bibliographystyle{abbrv}

\clearpage

\appendix

\section{Ziegler's proof of NP-hardness of MEI}

In his PhD-thesis~\cite{zieglerTh}, Ziegler showed that MEI is NP-hard (the 
corresponding decision problem is NP-complete) for general~$k$. Since this 
thesis is somewhat hard to obtain, we reproduce a slightly simplified version
of his proof. %However, we stress that the original proof is solely due to Ziegler.

\begin{theorem}[Ziegler \cite{zieglerTh}]
 Given a graph $G$, a set $F$ of unordered vertex-pairs, and an integer $b$, it is NP-complete
 to decide whether there is a planar drawing $D$ of\/ $G$ such that we can insert an 
 edge $vw$ for each vertex pair $\{v,w\}\in F$ into $D$ with overall at most $b$ crossings.
\end{theorem}
\begin{proof}
 NP-membership is trivial and it hence remains to show NP-hardness. We use a reduction from
 \begin{quote}
  \textsc{Fixed Linear Crossing Number (FLCN):}\\
  Given a graph $H=(V,E)$, a 1-to-1 function $f\colon V\to \{1,2,\ldots,|V|\}$, and an integer $\ell$.
  Does there exist an $f$-linear drawing of $H$ with at most $\ell$ crossings?  
 \end{quote}
 Thereby, an \emph{$f$-linear drawing} is one where all vertices are placed on a horizontal line, 
 each vertex $v$ at coordinate $f(v)$, and each edge is either drawn completely above or
 completely below that line. It was shown in~\cite{flcn} that {\sc Flcn} is NP-complete.
 
 Let $(H=(V,E),f,\ell)$ be an instance to FLCN. We will construct a corresponding MEI instance 
 $(G=(W,E^*),F,b)$ of size polynomial in $|V(H)+E(H)|$ which is a yes-instance for MEI if and only if $(H,f,\ell)$ is a yes-instance
 for FLCN. The key idea is to build a rigid graph $G$ that models the restrictions of 
 $f$-linear drawings, into which we then have to insert the original edges $E$.
 Observe that FLCN can only be hard if $\ell<|E|^2$.
 
 Label the vertices $V=\{v_1,\ldots,v_{n}\}$ such that $f(v_i)=i$ and $n=|V|$. 
 We may assume w.l.o.g. that $n\geq3$. We start with 
 constructing a graph $G'=(W,E')$ on $n+2$ vertices where $W:=V\cup\{w_a,w_b\}$
 and $E':=\{v_iv_{i+1} : 1\leq i<n\} \cup \{v_1w_a, v_nw_a, v_1w_b, v_nw_b, w_aw_b\}$.
 Observe that $G'$ is planar and (since its SPR-tree consists of one R- and one S-node) 
 allows only a unique embedding (and its mirror).
 We obtain $G$ from $G'$ by replacing each edge in $E'$ by $|E|^2$ parallel edges.  
 Up to ordering the multiple edges amongst its peers, $G$ still allows only a unique embedding.
 
 Now, set $b:=\ell$ and $F:=E$, i.e., we want to insert the edges of $H$ into planar $G$.
 We can assume w.l.o.g.\ that $E$ contains no edges $v_iv_i$ or $v_iv_{i+1}$ for any~$i$.
 $G$ has exactly four faces with more than two incident vertices: let $\varphi_1$ ($\varphi_2$) be 
 the face incident to exactly $\{w_a,w_b,v_1\}$ ($\{w_a,w_b,v_n\}$, respectively). Let $\varphi_a$ ($\varphi_b$)
 be the face incident to all of $V$ and $w_a$ ($w_b$, respectively). To go from one of these four faces to another,
 we would always have to cross at least $|E|^2$ edges (a parallel bunch), which is infeasible when asking for a 
 solution with at most $b=\ell<|E|^2$ crossings. No edge of $F=E$ will be placed in $\varphi_1$ or $\varphi_2$ as 
 they are only incident to one vertex of $V$. Hence each edge is either completely within $\varphi_a$ or $\varphi_b$,
 and the equivalence with being above or below the horizontal line of an $f$-linear drawing follows.\qed
\end{proof}

\end{document}